\PassOptionsToPackage{dvipsnames}{xcolor}
\documentclass[sigconf,preprint,nonacm]{acmart}

\setcopyright{none}

\usepackage{alltt}
\usepackage{amsmath,amsfonts}
\usepackage{algorithmic}
\usepackage{graphicx}
\usepackage{textcomp}
\usepackage{verbatim}
\usepackage{url}
\usepackage{array}
\newcolumntype{P}[1]{>{\centering\arraybackslash}p{#1}}
\usepackage{multirow}
\usepackage[normalem]{ulem}
\usepackage{cleveref}
\usepackage{xspace}
\usepackage{enumitem}
\usepackage{subcaption}
\usepackage{tabularx}
\usepackage{soul}
\usepackage{enumitem}
\usepackage{booktabs}
\usepackage{blindtext}
\usepackage{cleveref}
\usepackage{wasysym}

\usepackage{caption}
\usepackage{subcaption}
\usepackage[title]{appendix}

\usepackage[normalem]{ulem}
\useunder{\uline}{\ul}{}

\usepackage{pifont}

\def\BibTeX{{\rm B\kern-.05em{\sc i\kern-.025em b}\kern-.08em
    T\kern-.1667em\lower.7ex\hbox{E}\kern-.125emX}}

\newcommand{\code}[1]{\texttt{#1}}
\newcommand{\OpNameColor}[1]{\colorbox{CornflowerBlue!50}{#1}}
\newcommand{\ParamNameColor}[1]{\colorbox{SpringGreen!50}{#1}}
\newcommand{\ParamValueColor}[1]{\colorbox{Goldenrod!50}{#1}}
\newcommand{\MaskColor}[1]{\colorbox{Salmon!50}{#1}}

\graphicspath{ {figures/} }

\newcommand{\tool}{\textsc{LowCoder}\xspace}
\newcommand{\vptool}{\textsc{LowCoder}\textsubscript{VP}\xspace}
\newcommand{\nltool}{\textsc{LowCoder}\textsubscript{NL}\xspace}

\newcommand{\change}[1]{{\color{black} #1}}
\newcommand{\changeappendix}[1]{{\color{black} #1}}
\newcommand{\changenarrative}[1]{{\color{black} #1}}

\newcommand{\mlt}{\!\ll\!}

\newcolumntype{Y}{>{\centering\arraybackslash}X}

\begin{document}

\title{AI for Low-Code for AI}

\author[N. Rao]{Nikitha Rao}
\email{nikitharao@cmu.edu}
\affiliation{%
  \institution{Carnegie Mellon University}
  \state{United States}
}

\author[J. Tsay]{Jason Tsay}
\email{jason.tsay@ibm.com}
\affiliation{%
  \institution{IBM Research}
  \state{United States}
}

\author[K. Kate]{Kiran Kate}
\email{kakate@us.ibm.com}
\affiliation{%
  \institution{IBM Research}
  \state{United States}
}

\author[V. J. Hellendoorn]{Vincent J. Hellendoorn}
\email{vhellendoorn@cmu.edu}
\affiliation{%
  \institution{Carnegie Mellon University}
  \state{United States}
}

\author[M. Hirzel]{Martin Hirzel}
\email{hirzel@us.ibm.com}
\affiliation{%
  \institution{IBM Research}
  \state{United States}
}

\begin{abstract}
Low-code programming allows citizen developers to create programs with minimal coding effort, typically via visual (e.g.\ drag-and-drop) interfaces.
In parallel, recent AI-powered tools such as Copilot and ChatGPT generate programs from natural language instructions. 
We argue that these modalities are complementary: tools like ChatGPT greatly reduce the need to memorize large APIs but still require their users to read (and modify) programs, whereas visual tools abstract away most or all programming but struggle to provide easy access to large APIs.
At their intersection, we propose \tool, the first low-code tool for developing AI pipelines that supports both a visual programming interface (\vptool) and an AI-powered natural language interface (\nltool).
We leverage this tool to provide some of the first insights into whether and how these two modalities help programmers by conducting a user study. We task 20 developers with varying levels of AI expertise with implementing four ML pipelines using \tool, replacing the \nltool component with a simple keyword search in half the tasks.
Overall, we find that \tool is especially useful for (i)~Discoverability: using \nltool, participants discovered new operators in 75\% of the tasks, compared to just  32.5\% and 27.5\% using web search or scrolling through options respectively in the keyword-search condition, and (ii)~Iterative Composition: 82.5\% of tasks were successfully completed and many initial pipelines were further successfully improved. Qualitative analysis shows that AI helps users discover \emph{how} to implement constructs when they know \emph{what} to do, but still fails to support novices when they lack clarity on what they want to accomplish. Overall, our work highlights the benefits of combining the power of AI with low-code programming.
\end{abstract}

\maketitle
\section{Introduction}
Most AI development today involves Python programming with popular libraries such as scikit-learn (sklearn)~\cite{scikit-learn}.
Unfortunately, writing code, even in a language as high-level as Python, is hard for \emph{citizen developers}~\cite{ko_myers_aung_2004}\change{---people
who lack formal training in programming but nevertheless write programs as part of their everyday work. This is a fairly common situation for data scientists, among others.}
AI programming libraries also tend to be large and change regularly. Needing to remember hundreds of AI operators and their arguments slows down even professional developers.

\emph{Low-code programming}~\cite{sahay_et_al_2020} reduces the amount of textual code developers write by offering alternative programming interfaces. 
In recent years, it has been embraced by software vendors to both democratize software development and increase productivity.
Most low-code offerings for building AI pipelines currently favor visual programming
~\cite{berthold_et_al_2009,demsar_et_al_2004,hall_et_al_2009}. 
While visual programming helps users navigate complex pipelines, it poorly supports \emph{discoverability} of API components in large APIs due to the large range of options and limited screen space.
\change{In parallel, \emph{programming by natural language} (PBNL) has recently soared in popularity. Tools like Copilot~\cite{copilot} and ChatGPT~\cite{chatgpt} can generate code from natural language prompts in which users describe what they want to accomplish, which is especially helpful in ecosystems with large APIs. However, these tools still generate code, which can be complicated and hard to understand~\cite{copilotstudy2022}, especially without formal training in programming. }

\changenarrative{At the intersection of these two paradigms, we propose \tool, the first low-code tool to combine visual programming with PBNL.
We conjecture that the respective strengths of these two low-code techniques can compensate for each other's weaknesses. PBNL uses AI to help users retrieve and use programming constructs based on natural language queries. This does not always return correct programs, necessitating a way to help users understand and fix generated programs. Visual programming complements PBNL by providing a clear, unambiguous representation of the program that users can directly manipulate to experiment with alternatives.
Our goal is to help people who know \emph{what} they want to accomplish~(e.g., build a data processing pipeline) but face syntactic barriers from the programming language and library~(the \emph{how} part).
}

\tool's visual programming component, \vptool, lets users snap together visual blocks for AI operators into well-structured AI pipelines. It uses Blockly~\cite{pasternak_fenichel_marshall_2017} to provide a Scratch-like~\cite{resnick_et_al_2009} look-and-feel.
The PBNL component, \nltool, lets users enter natural language queries and predicts relevant operators, optionally configured with hyper-parameters. It uses a fine-tuned variant of the CodeT5 model~\cite{wang_et_al_2021} that we developed through experiments with a variety of neural models for program generation, ranging from training models from scratch to few-shot prompting large language models \cite{nijkamp_et_al_2022}. We further noticed that it is common in this domain for queries to mention at most a subset of hyper-parameters for each pipeline step, so we developed a novel task formulation tailored to this use case that improved learning outcomes.

\changenarrative{
We leverage \tool to provide some of the first insights into both how and when low-code programming and PBNL help developers with various degrees of expertise.
We conduct a user study with 20~participants with varying levels of AI expertise using \tool to complete four tasks, half of which with, and half without the help of the AI-powered component \nltool.
Overall, we find that the combination of visual programming along with the natural language interface helped both novice and non-novice users to successfully compose \change{pipelines (85\% of tasks)} and then further refine their pipelines \change{(72.5\% of tasks)} during the study when using \nltool.
Additionally, \nltool helped users discover previously-unknown operators in 75\% of the tasks compared to just 32.5\% using other methods like web search when \nltool was not available. In addition, despite being trained on a different dataset, \nltool accurately answered real user queries.
}
In summary, his paper makes three main contributions:

\begin{itemize}[leftmargin=6mm,nolistsep]
\item[(i)] Low-Code for AI: \change{We introduce \tool, a new low-code tool that combines visual programming and PBNL to help develop AI pipelines.}

  \item[(ii)] AI for Low-Code: \change{We benchmark various AI models and develop a novel task formulation to develop an AI powered natural language interface to \tool.}
  
  \item[(iii)] User Study: \changenarrative{We analyze the trade-offs between the two modalities and study the effects of using AI for low-code programming through a user study involving 20 participants with varying levels of AI expertise using \tool.}
\end{itemize}

\section{Related Work}\label{sec:related}

\textbf{\textcolor{gray}{AI for} \underline{Low-code} \textcolor{gray}{for AI}:}
In adopting a visual programming approach to low-code, we follow a long tradition~\cite{boshernitsan_downes_2004}. We were particularly inspired by Scratch, a popular visual programming environment for children that uses lego-like connected blocks~\cite{resnick_et_al_2009}. Our other inspiration came from projectional editors, where the visual programming interface is a projection, or \emph{view}, over an internal domain-specific language~(DSL)~\cite{voelter_lisson_2014}. Our implementation uses Blockly, a meta-tool for creating block-based visual programming tools~\cite{pasternak_fenichel_marshall_2017}, and Lale, a DSL for AI pipelines~\cite{baudart_et_al_2021}.

\begin{figure}
    \centering
    \vspace{-2mm}
    \includegraphics[width=0.9\linewidth,trim={17cm 12.5cm 17cm 12cm},clip]{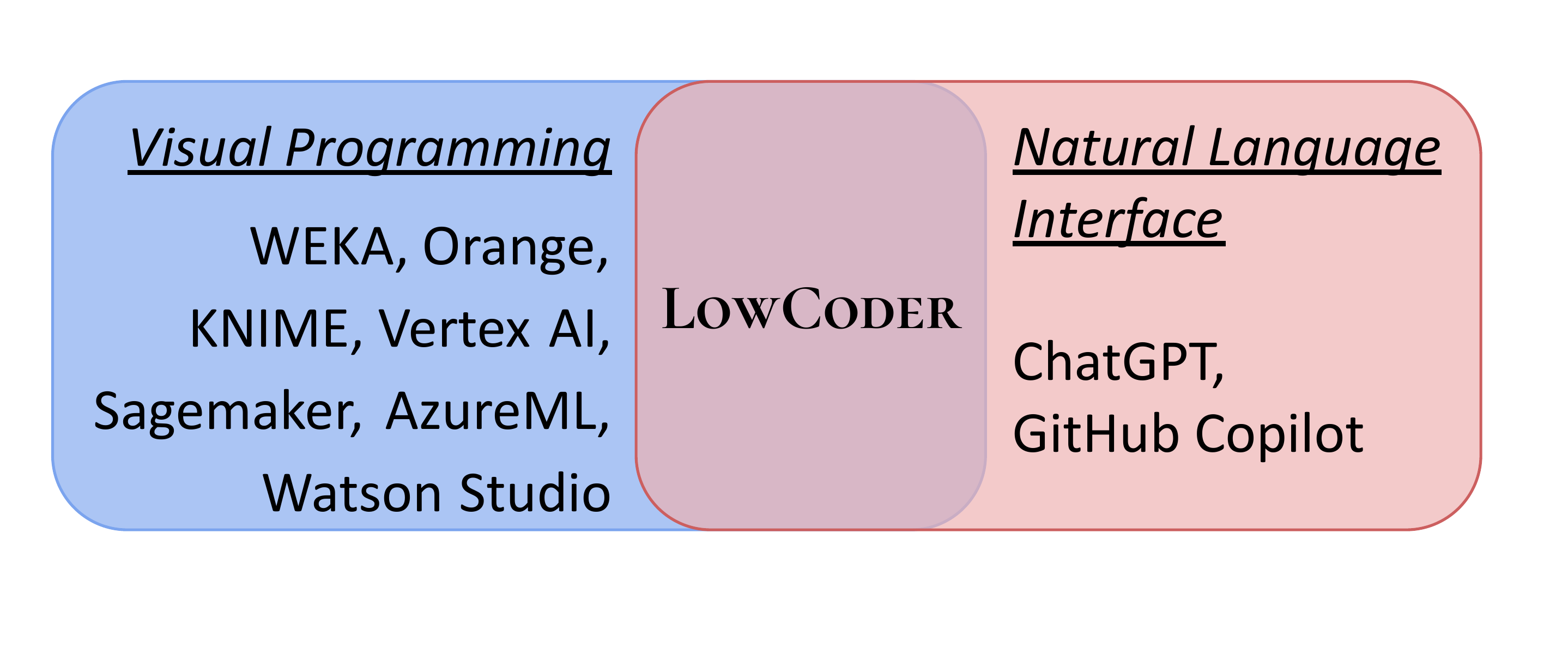}
    \vspace{-8mm}
    \caption{Relationship between \tool and other low-code for AI tools.}
    \label{fig:comparison}
    \vspace{-4mm}

\end{figure}

\textbf{\textcolor{gray}{AI for} \underline{Low-code for AI}:} Most low-code interfaces for programming AI pipelines use visual programming. Examples include WEKA~\cite{hall_et_al_2009}, Orange~\cite{demsar_et_al_2004}, and KNIME~\cite{berthold_et_al_2009}. Each has a palette of operators that can be dragged onto a canvas, where they can be connected into a boxes-and-arrows style diagram. 
\change{Commercial low-code visual interfaces follow the same approach, such as Vertex AI, Sagemaker, AzureML, and Watson Studio.
A related approach for low-code AI pipeline development is automated machine learning (AutoML) which is also used by many of the same commercial AI interfaces mentioned earlier. These tools tend to have a black-box approach where the user has little control over the AutoML search and may not even see the resulting pipeline. AutoML libraries such as auto-sklearn~\cite{feurer2015}, TPOT~\cite{pmlr-v64-olson_tpot_2016}, and hyperopt~\cite{bergstra2013} provide a Python interface, which is intended for 
\change{textual code} development. 
There are also natural-language interfaces for professional developers based on large language models such as GitHub Copilot which uses Codex~\cite{chen_et_al_2021} and ChatGPT. Since these support APIs for which there is sufficient publicly available code to use as training data, they cover popular machine learning libraries such as sklearn. 
The main difference between these low-code for AI tools and our paper is that we combine the ease-of-use of visual programming with a natural language interface to help users discover and configure operators and, inspired by Scratch~\cite{resnick_et_al_2009}, our tool encourages liveness~\cite{tanimoto_2013} through immediate user feedback for each user input into the system. This contrasts with most tools that require explicit training and scoring steps for feedback.
\Cref{fig:comparison} summarizes the relationship between \tool and other low-code for AI tools.}

\textbf{\underline{AI for Low-code} \textcolor{gray}{for AI}:} The most prominent AI technique for low-code is programming by natural language~(PBNL).
When Androutsopoulos et al.\ surveyed natural language interfaces to databases in 1995, it was already a well-established field~\cite{androutsopoulos_ritchie_thanisch_1995}. Desai et al.\ treat PBNL as a program synthesis problem targeting a DSL designed for the purpose~\cite{desai_et_al_2016}. The Overnight paper addresses the problem of missing training data for PBNL interfaces by crowd-sourcing~\cite{wang_berant_liang_2015}. And SwaggerBot lets users extend and customize a chatbot from within the chatbot itself~\cite{vaziri_et_al_2017}. Unlike these works, our paper uses large language models for PBNL, uses PBNL for creating AI pipelines, and 
integrates with a visual programming interface.

\textbf{Combining low-code techniques:}
Our work combines visual programming with PBNL. In a similar vein, Rousillon combines visual programming with programming by demonstration~\cite{chasins_mueller_bodik_2018} and Pumice combines programming by demonstration with PBNL~\cite{li_et_al_2019}. Like Rousillon and Pumice, our goal in combining techniques is to use strengths of each technique to mitigate weaknesses in the other. However, unlike Rousillon and Pumice, we choose different techniques to combine, and we target a different domain, namely AI.

\change{\textbf{User studies on AI tools:} There are a few studies that aim to evaluate whether developers perform better on programming tasks when working with AI tools. 

Vaithilingam et al.\ had developers use GitHub Copilot on three programming tasks and found that while neither task success rate nor completion time improved while using Copilot, developers preferred using it compared to the standard code completion~\cite{copilotstudy2022}.
Similarly, Xu et al.\ had developers perform several programming tasks with and without the use of a natural language to code generation model and found no significant differences with regards to code quality, task completion time and program correctness~\cite{nl2code_ide_study}. Wang et al. interviewed several data scientists to better understand their perceptions of automated AI and found that they had mixed feelings~\cite{datascientist_study}. However, nearly all of them felt that the future of data science involved collaboration between humans and AI systems. Unlike other work which tends to focus on how AI supports software development by experienced developers, our paper focuses on AI tools in the context of low-code systems where developers have varying expertise levels in both building software and AI. }

\section{Low-Code for AI}
\label{sec:lc4ai}

This work explores the intersection of low-code and AI \change{in an effort to understand the benefits and limitations of using low-code for AI}. We accomplish this by implementing and studying \tool, a prototype low-code tool for building AI pipelines with sklearn operators for tabular data that includes both visual programming~(VP) and natural language~(NL) modalities, which complement each other by mitigating the limitations of either modality separately. Building this tool provided us with the opportunity to examine the impact of both modalities on users. Figure~\ref{fig:tool-interface} highlights the main features and inputs of \tool.

To support multiple low-code modalities, we follow the lead of projectional code editors~\cite{voelter_lisson_2014} by adopting the model-view-controller pattern. 
Specifically, we treat visual programming as a read-write view, PBNL as a write-only view, and let users inspect data in a read-only view.
The tool keeps these three views in sync by representing the program in a domain-specific language~(DSL).
The domain for the DSL is AI pipelines. A corresponding, practical desideratum is that the DSL is compatible with sklearn~\cite{scikit-learn}, the most popular library for building AI pipelines, and is a subset of the Python language, in which sklearn is implemented, which also enables us to use AI models pretrained on Python code.
The open-source Lale library~\cite{baudart_et_al_2021} satisfies these requirements, and in addition, describes hyper-parameters in JSON schema format~\cite{pezoa_et_al_2016}, which our tool also uses.
\change{The current version of our tool supports 143 sklearn operators.}
\tool uses a client-server architecture with a Python Flask back-end server and front-end based on the Blockly~\cite{pasternak_fenichel_marshall_2017} meta-tool for creating block-based visual programming tools.
The front-end converts the block-based representation to Lale which is then sent to the back-end.
The back-end validates the given Lale pipeline using internal schemas, then evaluates the pipeline against a given dataset.
The results of this evaluation (including any error messages) are returned to the front-end and presented to the user.

\subsection{Visual Programming Interface}

\vptool is our block-based visual programming interface for composing and modifying AI pipelines.
One goal that this tool shares with other block-based visual tools such as Scratch~\cite{resnick_et_al_2009} is to encourage a \change{highly interactive} experience. The block visual metaphor allows for blocks that correspond to sklearn operators to be snapped together to form an AI pipeline. The shape of the blocks suggest how operators can connect. Their color indicates how they affect data: red for operators that transform data~(with a \textit{transform()} method) and purple for other operators 
that make predictions, such as classifiers and regressors~(with a \textit{predict()} method).

\begin{figure}[t]
    \centering
    \vspace{2mm}
    \includegraphics[width=\linewidth]{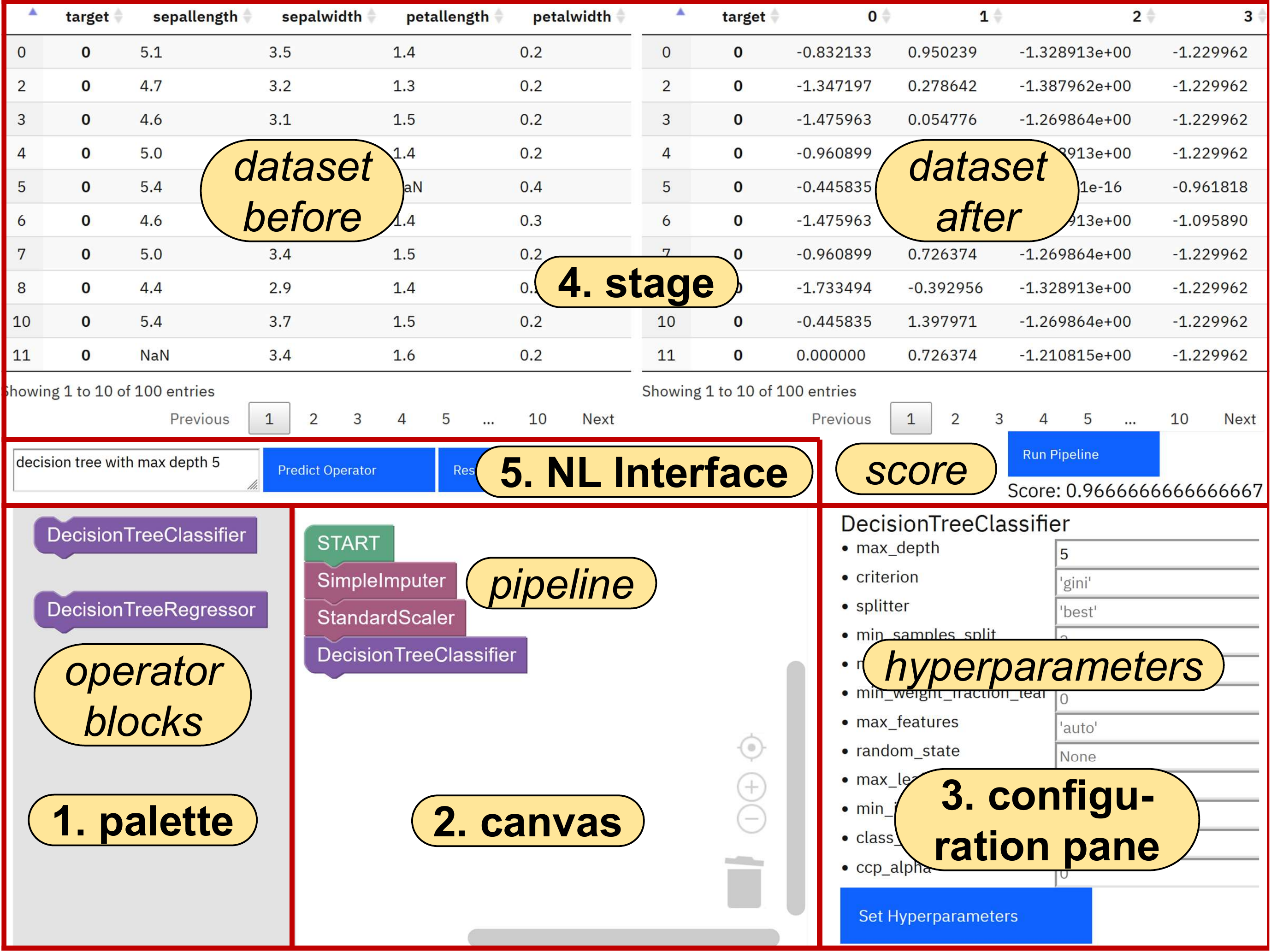}
    \vspace{-7mm}
    \caption{\tool interface with labeled components, described in the text.}
    \label{fig:tool-interface}
    \vspace{-4mm}    
\end{figure}

A \textit{palette~(1)} on the left side of the interface contains all of the available operator blocks. Blocks can be dragged-and-dropped from the palette to the \textit{canvas~(2)}. For ease of execution, our tool only allows for one valid pipeline at a time, so blocks must be attached downstream of the pre-defined \textit{Start} block to be considered part of the active pipeline. Figure~\ref{fig:tool-interface} shows an example of blocks defining a pipeline where the \texttt{SimpleImputer}, \texttt{StandardScaler}, and \texttt{DecisionTreeClassifier} blocks are connected to the \textit{Start} block and each other. Input data are transformed by the first two operators (\texttt{SimpleImputer} and \texttt{StandardScaler}) and then sent to \texttt{DecisionTreeClassifier} for training and then scoring. Blocks not attached to the \textit{Start} block are disabled but can be left on the canvas without affecting the execution of the active pipeline. 
Selected operator blocks also display a \textit{hyper-parameter configuration pane~(3)} on the right. The pane lists each hyper-parameter for an operator along with a description (when hovering over the hyper-parameter name) and default values along with input boxes to modify each hyper-parameter.

\change{Our tool provides a \emph{stage~(4)} with \textit{Before} and \textit{After} tables to give immediate feedback with every input on how the current pipeline affects the given dataset. 
When a tabular dataset is loaded, the \textit{Before} table displays its target column on the left and feature columns on the right.
When a pipeline that transforms input data is executed, the \textit{After} table shows the results of the transformations.}
At any time, a pipeline can be executed on the given dataset by pressing the ``Run Pipeline'' button. Executing a pipeline will attempt to train the given pipeline on the training portion of the given dataset and then return a preview of all data transformations on the training data in a second table. 
For instance, in the example shown in \Cref{fig:tool-interface}, executing the pipeline with \texttt{SimpleImputer} and \texttt{StandardScaler} transforms data from the \textit{Before} table by imputing missing values and standardizing all feature values in the \textit{After} table.
If training is successful, then the trained pipeline is scored against the test set and the score (usually accuracy) is displayed.
\vptool also encourages liveness~\cite{tanimoto_2013} by executing the pipeline when either the active pipeline is modified or hyper-parameters are configured.
For example, adding a \texttt{PCA} operator and setting the \texttt{n\_components} hyper-parameter to 2 for the prior example will reduce the feature columns in the \textit{After} table to~2. 
This gives the user immediate feedback on the effect that pipeline changes have on the dataset \change{without requiring separate training or scoring steps. This liveness encourages a high degree of interactivity~\cite{resnick_et_al_2009}.}

\subsection{Natural Language Interface}
A potential weakness of visual low code tools is that users have trouble discovering the right components to use~\cite{ko_myers_aung_2004}. For instance, the palette of \vptool contains more than a hundred operator blocks. Rather than requiring users to know the exact name of the operator or scroll through so many operators, we provide \nltool, which allows users to describe a desired operation in the \textit{NL interface~(5)} text box and press the ``Predict Pipeline'' button. The tool then infers relevant operator(s) and any applicable hyper-parameters using an underlying natural language to code translation model and automatically adds the most relevant operator to the end of the pipeline. The palette is also filtered to only display any relevant operator(s) such as in \Cref{fig:tool-interface}. Pressing the ``Reset Palette'' button will undo filtering (so the palette shows all available operators again) without clearing the active pipeline or canvas. Depending on the NL search, the automatically added operator may either have hyper-parameters explicitly defined or potentially relevant hyper-parameters highlighted. As an example, the NL search \textit{``PCA with 2 components''} will automatically add the \texttt{PCA} operator where the \texttt{n\_components} hyper-parameter is set to 2 and may highlight other hyper-parameters such as \texttt{random\_state} for the user to consider setting. 
\Cref{sec:ai4lc} describes the design and implementation of this model in detail. A potential weakness of natural language low-code tools is that the generated programs can be incorrect, due to a lack of clarity, or ambiguity, in the query, or a lack of context for the model providing inferences~\cite{androutsopoulos_ritchie_thanisch_1995}. In comparison, visual inputs and representations are unambiguous~\cite{hirzel_2022}, requiring no probabilistic interpretation, so users can easily understand and manipulate the results returned by \nltool.

To ground our evaluation of \nltool, we also provide a version of the tool without a trained language model to users in our study (described in \Cref{sec:evaluation}). In this setting, the \textit{NL interface (5)} text box becomes a simple substring keyword search that matches the query against operator names. For example, inputting \textit{``classifier''} filters the palette to only display sklearn operators that contain \textit{`classifier'} in the name such as \texttt{RandomForestClassifier} (but notably not all classifiers such as \texttt{SVC}). 

\section{AI for Low-Code}
\label{sec:ai4lc}

This section discusses the AI that went into \nltool.

\subsection{Data Collection}
Our goal is to make a large API accessible through a low-code tool by allowing users to describe \emph{what} they want to do when they do not know \emph{how}. More specifically, we want to enable users to build sklearn pipelines in a low-code setting, using a natural language interface that can be used as an \emph{intelligent search} tool. This problem can be solved using language models that can be trained to translate a natural language query into the corresponding line of code~\cite{Codebert}. However, such models heavily rely on data to learn such behaviour and would need to be trained on an aligned dataset of natural language queries and the corresponding sklearn line(s) of code demonstrating how a user would want to use such an intelligent search tool. 
Naturally, we cannot collect such a dataset without this tool, creating a circular dependency.
To overcome this challenge, we curate a \emph{proxy dataset} using 140K Python Kaggle notebooks that were collected as part of the Google AI4Code challenge.\footnote{https://www.kaggle.com/competitions/AI4Code} From these notebooks, we extracted aligned Natural Language (NL) \& Code cells related to machine learning and data science tasks. While the distribution of the NL in the markdown cells is not completely representative of the NL queries that users would enter in the low-code setting, they provide the model with a broad range of such examples.
Results in \Cref{sec:model-results} show that this is indeed effective.

\subsection{Data Preprocessing}
\label{sec:data-preprocess}
We first filter out notebooks that do not contain any sklearn code. This leaves 84,783 notebooks -- evidently, many notebooks involve sklearn. 
We further filter out notebooks with non-English descriptions in all of the markdown cells, resulting in 59,569 notebooks.
We then create a proxy dataset by extracting all code cells containing sklearn code and pairing these with their preceding NL cell to get a total of 211,916 aligned NL-code pairs. We remove any duplicate NL-code pairs, leaving 102,750 unique pairs. For each code cell, we then extract the line(s) of code corresponding to an sklearn operation invocation statement. 
 
We discard any code cells that do not include sklearn operation invocation statements but include other sklearn code

leaving a final total of 79,372 NL-Code pairs. We separate these into train/validation/test splits resulting in 64,779 train samples, 7,242 validation samples, and 7,351 test samples. \changeappendix{See Section B in the appendix for more details.}

\subsection{Tasks}

\begin{table*}[th]
    
  \begin{center}
  \small
    \caption{\label{tab:task-types}Task formulations highlighting the code components: \MaskColor{mask}, \OpNameColor{operator name}, \ParamNameColor{hyper-parameter name}, \ParamValueColor{hyper-parameter value}. The Hybrid Operator Invocation setting does not mask `balanced' as it appears in the query.}
    \vspace{-4mm}
    \begin{tabular}{l|l}

    {\textbf{Task Formulation}} & \textbf{Code for the NL query: \emph{Random forest with balanced class weight}} \\
    \hline
    Operator Name & \code{\OpNameColor{RandomForestClassifier}}\\
    Complete Operator Invocation & \code{\OpNameColor{RandomForestClassifier}(\ParamNameColor{n\_estimators}=\ParamValueColor{100}, \ParamNameColor{class\_weight}=\ParamValueColor{'balanced'})} \\ 
    Masked Operator Invocation & \code{\OpNameColor{RandomForestClassifier}(\ParamNameColor{n\_estimators}=\MaskColor{MASK}, \ParamNameColor{class\_weight}=\MaskColor{MASK})} \\ 
    Hybrid Operator Invocation & \code{\OpNameColor{RandomForestClassifier}(\ParamNameColor{n\_estimators}=\MaskColor{MASK}, \ParamNameColor{class\_weight}=\ParamValueColor{'balanced'})}\\

    \end{tabular}
      \vspace{-2mm}
  \end{center}
\end{table*}

\label{sec:tasks}
Given the NL query, our model aims to generate a line of sklearn code corresponding to an operation invocation that can be used to build the next step of the pipeline.
We consider a range of formulations of the task with different levels of details, as illustrated in \Cref{tab:task-types}. \changeappendix{Additional examples can be found in Section A of the appendix.}

\subsubsection{Operator Name Generation}
The simplest task is generating only the operator name from the NL query. This alone can significantly help a developer with navigating the extensive sklearn API. We process the aligned dataset to map the query to the name(s) of operator(s) invoked in the code cell, discarding any other information such as hyper-parameters.
 
\subsubsection{Complete Operator Invocation Generation} 
At the other extreme, we task the model with synthesizing the complete operation invocation statement, including all the hyper-parameter names and values. Preliminary results (discussed in \Cref{sec:task-results}) show that the model often makes up arbitrary hyper-parameter values, resulting in lines of code that can rarely be used directly by developers.

\subsubsection{Masked Operator Invocation Generation} 
In this scenario, we mask out all the hyper-parameter values from the invocation statement, keeping only their names.
The goal of this formulation is to ensure that the model learns to predict the specific invocation signature, even if it is unaware of the values to provide for the hyper-parameters.

\subsubsection{Hybrid Operator Invocation Generation (HOI)}
Manual inspection of the NL-code pairs revealed that the queries sometimes explicitly describe a subset of the hyper-parameter names and values to be used in the code. When this is the case, the model has the necessary context to predict at least those hyper-parameter values. Supporting this form of querying enables users to express the most salient hyper-parameters up-front. Therefore, we formulated a new hybrid task, where we keep the hyper-parameter values if they are explicitly stated in the NL query and mask them otherwise. This gives the model an opportunity to learn the hyper-parameter names and values if they are explicitly stated in the description, and unburdens it from making up values that it lacks the context to predict by allowing it to generate placeholders~(masks) for them.

\noindent\textbf{Evaluation:} To evaluate the feasibility of predicting code using the different task formulations, we train a simple sequence-to-sequence model (detailed in Section~\ref{sec:transformer}) and compare the results for the various training tasks in Section~\ref{sec:task-results}. We find HOI to be the most accurate/reliable formulation for our setting. We therefore proceed to use this task formulation for training the models.

\subsection{Modeling}
\label{sec:modeling}
\begin{figure}[b]
    \centering
    \vspace{-2mm}
    \includegraphics[width=\linewidth]{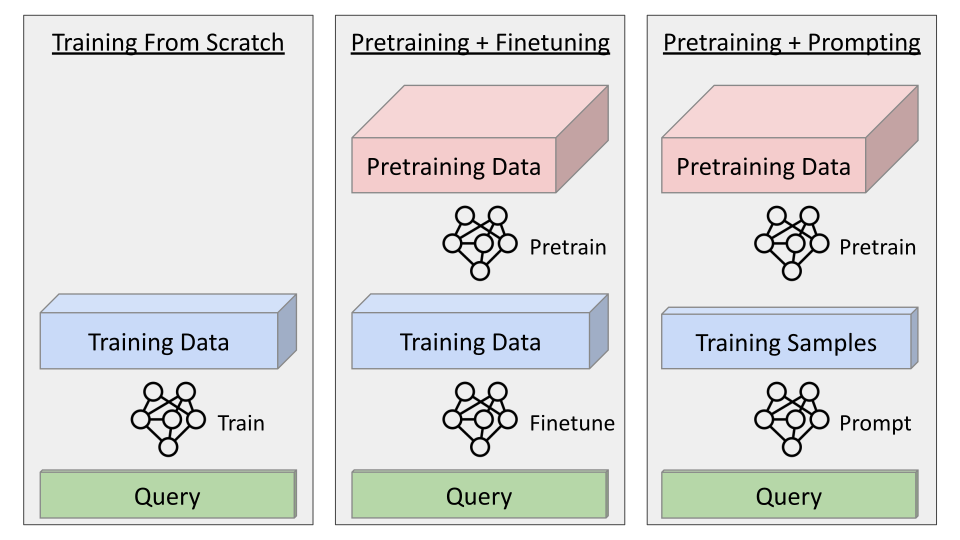}
    \vspace{-8mm}
    \caption{Overview of the ``trifecta" of training approaches used in contemporary deep learning: smaller models are directly trained from scratch on downstream task data; medium sized models (100M-1B parameters) are pretrained with a generic training signal and then fine-tuned on task data; large models ($>$1B parameters) are only pretrained on very large datasets and are prompted with examples from the training data as demonstration followed by the query.}
    \label{fig:modeling-trifecta}
    \vspace{-5mm}
\end{figure}

All tasks from \Cref{sec:tasks} are sequence-to-sequence tasks.
We compare and contrast three different deep learning paradigms for this type of task, illustrated in \Cref{fig:modeling-trifecta}: \change{1) train a standard sequence-to-sequence transformer \emph{from scratch}, 2) fine-tune (calibrate) a pretrained ``medium'' sized model, 3) query a Large Language Model (LLM) by means of few-shot prompting~\cite{Radford19}.}
We elaborate on these models below. Note that we use top-k sampling for our top-5 results. (A comparison of results with other decoding strategies can be found in Section 3 and 4 in the supplementary material). 

\subsubsection{Transformer (from scratch)}
\label{sec:transformer}
We train a sequence-to-sequence Transformer model \cite{Vaswani_17} with randomly initialized parameters on the training data.
\change{Our relatively small dataset of ca.\ 70K training samples limits the size of a model that can be trained in this manner.}
We use a standard model size, with 6 encoder and decoder layers and 512-dimensional attention across 8 attention heads and a batch size of 32 sequences with up to 512 tokens each. We use a sentence piece tokenizer (trained on Python code) with a vocabulary size of 50K tokens. The model uses an encoder-decoder architecture that jointly learns to encode (extract a representation of) the natural language sequence and decode (generate) the corresponding sklearn operator sequences.

\subsubsection{Fine-tuning CodeT5}
CodeT5 is a pretrained encoder-decoder transformer model \cite{wang_et_al_2021} that has shown strong results when fine-tuned (calibrated) on various code understanding and generation tasks~\cite{Lu2021CodeXGLUEAM}.
CodeT5 was pretrained on a corpus of six programming languages from the CodeSearchNet dataset~\cite{husain2019codesearchnet} and fine-tuned on several tasks from the CodeXGLUE benchmark\cite{Lu2021CodeXGLUEAM} in a multi-task learning setting, where the task type is prepended to the input string to inform the model of the task.
We fine-tune CodeT5 on the HOI generation task by adding the `Generate Python' prefix to all NL queries. We experiment with different size CodeT5 models: codet5-small (60M parameters), base (220M) and large (770M). 

\subsubsection{Few-Shot Learning With CodeGen}
Lastly, we explore large language models (LLMs) that are known to perform well in a task-agnostic few-shot setting \cite{brown2020}. More specifically, we look at CodeGen, a family of LLMs that are based on standard transformer-based autoregressive language modeling \cite{nijkamp_et_al_2022}. Pretrained CodeGen models are available in a broad range of sizes, including 350M, 2.7B, 6.1B and 16.1B parameters. These were all trained on three different datasets, starting with a large, predominantly English corpus, followed by a multi-lingual programming language corpus, and concluding with fine-tuning on just Python data, which we use in this work. The largest model trained this way was shown to be competitive with Codex \cite{chen_et_al_2021} on a Python benchmark \cite{nijkamp_et_al_2022}.

\begin{figure}[t]
    \centering
    \vspace{1mm}
    \includegraphics[width=\linewidth, trim ={11cm 8.5cm 6cm 9cm}, clip]{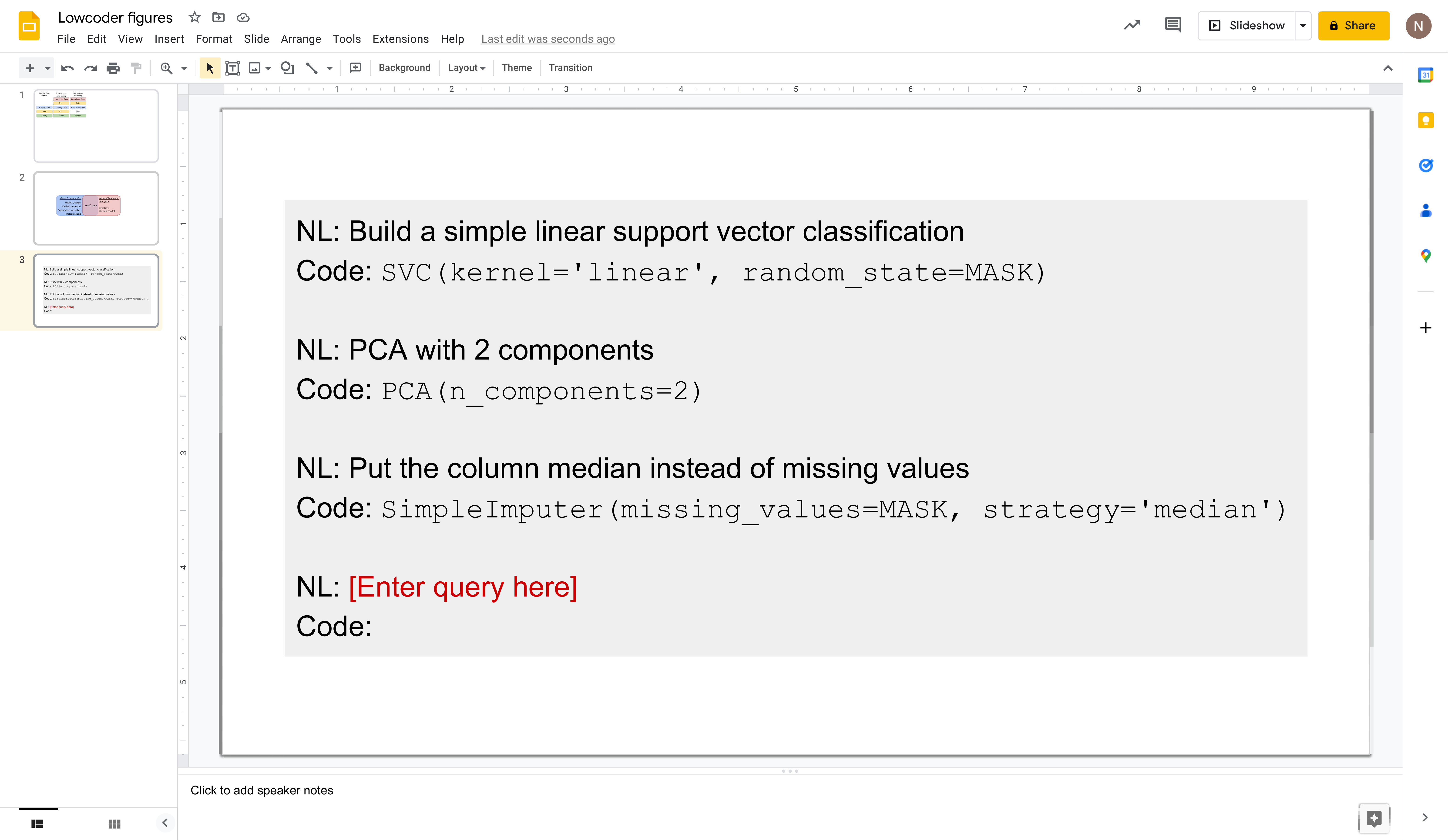}
    \vspace{-2mm}
    \caption{Example of a few (3) shot prompting template for querying a large language model in our study.}
    \label{fig:prompts}
   \vspace{-2mm}
\end{figure}

Models at this scale are expensive to fine-tune and are instead commonly used for inference by means of ``few-shot prompting"~\cite{Radford19}. 
LLMs are remarkably capable of providing high-quality completions given an expanded prompt containing examples demonstrating the task~\cite{brown2020}. 
We prompt our model with 5 such NL-code examples. \Cref{fig:prompts} illustrates an example prompt with 3 such pairs. The model learns from the examples in the prompt and completes the sequence task which results in generating the HOI code.

\section{Evaluation}\label{sec:evaluation}

This section describes the evaluations for the AI modeling that enables \nltool along with the user studies 
\change{that we conducted to analyze the benefits and challenges of using low-code for developing AI pipelines using \tool.}

\subsection{Modeling}

\subsubsection{Experimental Setup}

All of our models are implemented using PyTorch transformers and the HuggingFace interface. We use the latest checkpoints of the CodeT5~\cite{wang_et_al_2021} and CodeGen~\cite{nijkamp_et_al_2022} models.
Our models were trained on a single machine with multiple 48 GB NVIDIA Quadro RTX 8000 GPUs until they reached convergence on the validation loss.
We clip input and output sequence lengths to 512 tokens, but reduce the latter to 64 when using the model in \tool to reduce inference time. \change{We find in additional experiments that} since few predictions are longer than this threshold, this incurs no significant decrease in accuracy, but speeds up inference by 34\%.
We use a batch size of 32 for training and fine-tuning all of our Transformer and CodeT5 models, except for CodeT5-large, for which we used a batch size of 64 to improve stability during training. 

\subsubsection{Test Datasets}
\label{sec:test-data}
To ensure a well-rounded evaluation, we look at two different test datasets. 

\noindent\textbf{(i) Test data (from notebooks)} - We use the NL-code pairs from the Kaggle notebooks we created in~\Cref{sec:data-preprocess} containing 7,351 samples. These are noisy -- some samples contain vague and underspecified Natural Language~(NL) queries, such as - \textit{``Data preprocessing''}, \textit{``Build a model''}, \textit{``Using a clustering model''}. Others contain multiple operator invocation statements corresponding to a single NL query, even though the NL description only mentions one of them, e.g., \textit{``Model \# 2 - Decision Trees''} corresponds to \texttt{DecisionTreeClassifier()} and \texttt{confusion\_matrix(y\_true, y\_pred)}. Furthermore, these samples were collected from Kaggle notebooks, so the distribution of the NL queries collected from the markdown cells are not necessarily representative of NL queries that real users may enter into \nltool.

\noindent\textbf{(ii) Real user data} - We log all the NL queries that users searched for in \tool during the user studies along with the list of operators that the model returned. This gives us a more accurate distribution of NL queries that developers use to search for operators in \nltool. We obtained a total of 218 samples in this way, which we then manually annotated to check whether (i) the predictions were accurate, that is, if the operators in any of the predictions matches the inferred intent in the query and (ii) the NL query was clear, with an inter-rater agreement of 97.7\% and a negotiated agreement~\cite{garrison2006revisiting} of 100\%. \changeappendix{(See Section E in appendix for details on annotation guidelines.)}

\subsubsection{Test Metrics}
We use both greedy (top-1) and top-K (top-5) decoding \changeappendix{}{(see Section C in appendix)} when generating the operator invocation sequences for each NL query. We evaluate the models' ability to generate just the operator name as well as the entire operator invocation (including all the hyper-parameter names and values) based on the hybrid formulation.

\subsubsection{Task Comparison}
\label{sec:task-results}
\begin{figure}[t]
    \centering
    \vspace{-2mm}
    \includegraphics[width=\linewidth]{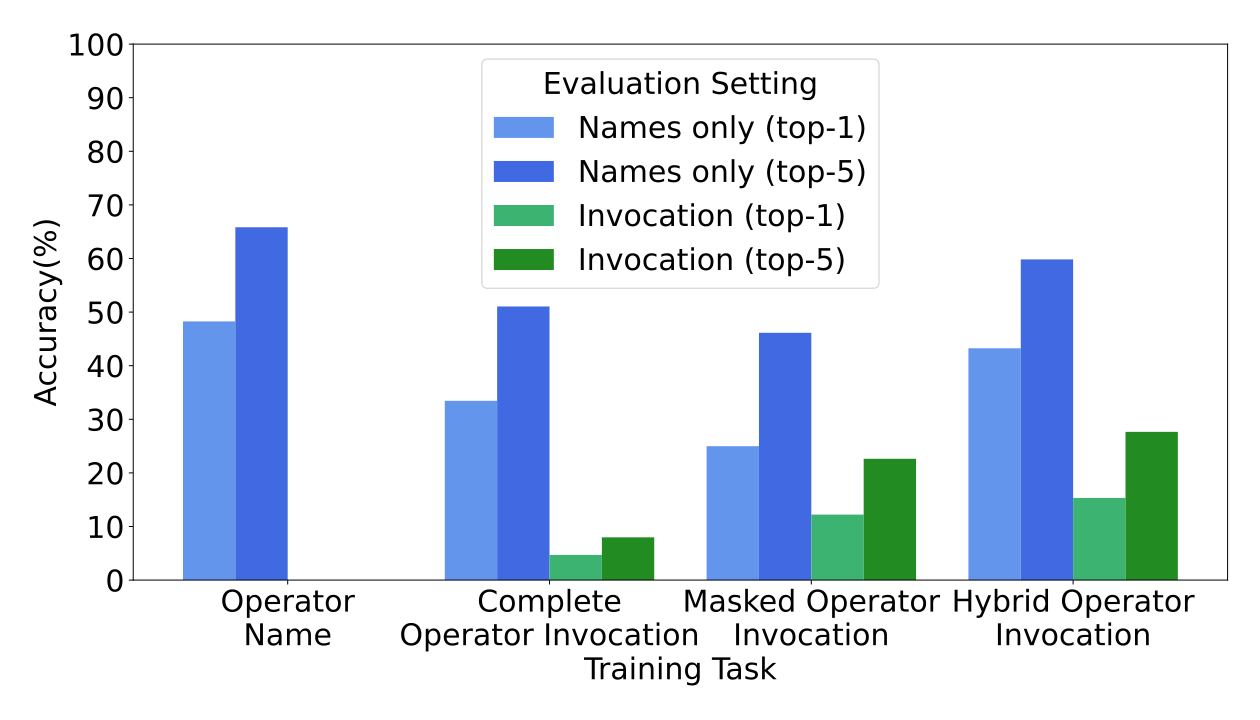}
    \vspace{-10mm}
    \caption{Accuracy of Transformer models trained from scratch on various task formulations.  `Invocation' test results refer to the specific invocation formulation of the training task, while `Names only' just considers whether the generated code starts with the correct operator name. Only the Hybrid Operator Invocation setting yields useful quality on both tasks.}
    \label{fig:task-results}
\vspace{-4mm}
    
\end{figure}

We first train a series of randomly initialized 6-layer Transformer models from scratch on each task formulation from \Cref{sec:tasks}.
We compare the model's ability to correctly generate the operator name and the operator invocation based on the formulation corresponding to the training task using top-1 and top-5 accuracy as shown in \Cref{fig:task-results}.
We find that the hybrid formulation of the operation invocation task, while challenging, is indeed feasible and allowed the model to achieve reasonably strong performance when generating the entire operation invocation statement. Contrary to the other task formulations, a model trained with the HOI signal also achieved comparable performance to the model trained solely on operator names when evaluated purely on operator name prediction (ignoring the generated hyper-parameter string). These results highlight that the hybrid representation helps the model learn by unburdening it from inferring values that it lacks the context to predict.

\begin{figure}
\centerline{\includegraphics[width=.8\columnwidth]{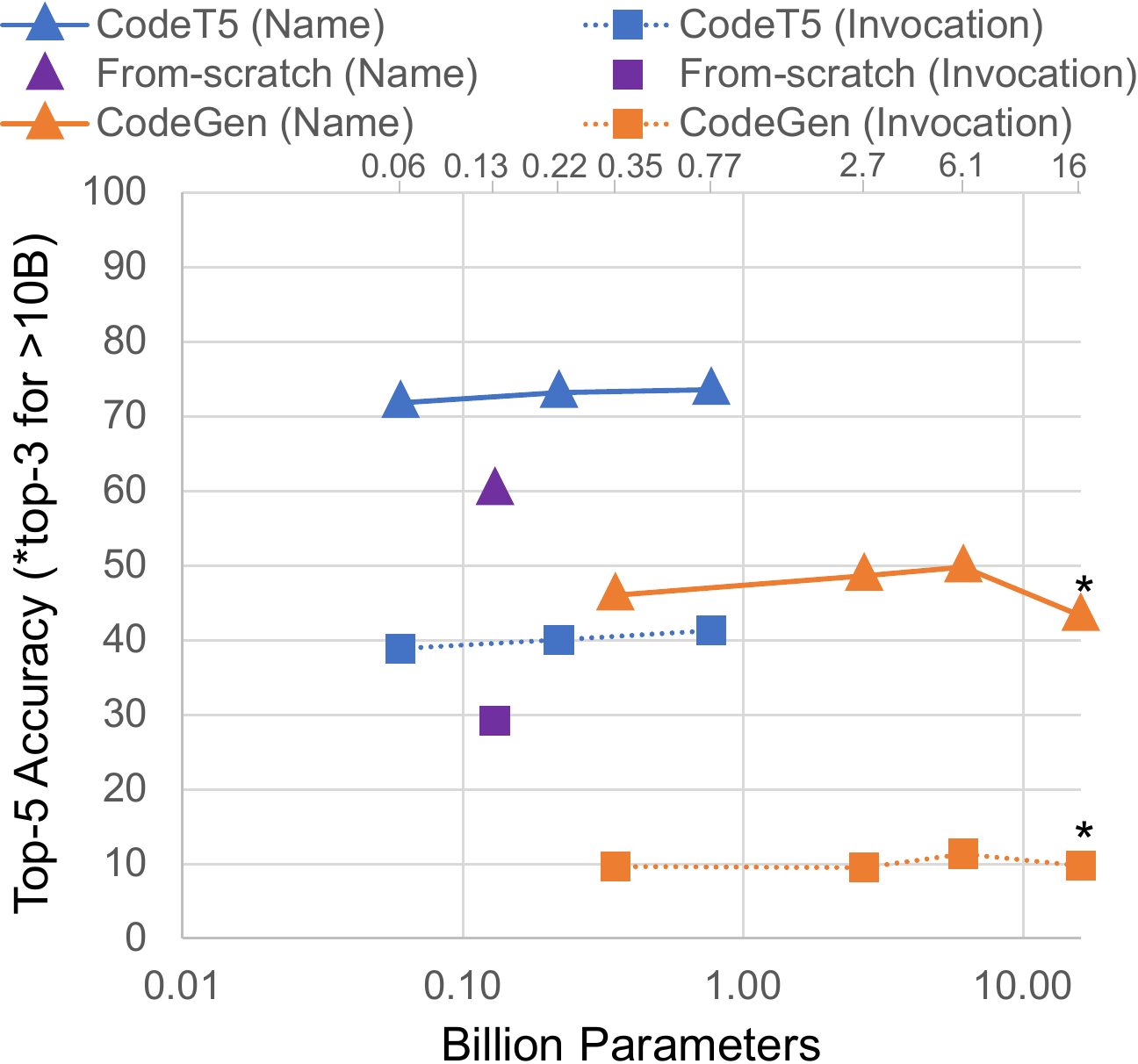}}
\vspace{-2mm}
\caption{\label{fig:model-results}Accuracy vs.\ model size based on top-5 sampling. (*The 16B CodeGen uses top-3 due to memory constraints.) \change{We compare the three modeling paradigms, namely training transformer from scratch, finetuning CodeT5, and fewshot prompting CodeGen, on both Operator Name generation and Hybrid Operator Invocation generation.}}
\vspace{-2mm}
\end{figure}

\subsubsection{Model Comparison}
\label{sec:model-results} 
We next evaluate the performance of the trifecta of modeling strategies from \Cref{sec:modeling} on the task of Hybrid Operation Invocation (HOI) generation. We benchmark across different model sizes and compare the performance for both operator name and operator invocation generation using top-5 accuracy in \Cref{fig:model-results}. \changeappendix{(See Section D in the appendix for additional results and ablation studies.)}
The results show that the 0.77B parameter fine-tuned CodeT5 is the best performing model with an accuracy of 73.57\% and 41.27\% on the test data for the operation name and operation invocation generation respectively. The 0.22B parameter fine-tuned CodeT5 model has comparable performance, but its inference time is approximately 2--3 seconds faster than the 0.77B fine-tuned CodeT5 model, making it more desirable for integration with the tool.

\subsubsection{Performance in Practice}
Up to this point, all our evaluations have been based on the proxy dataset from Kaggle. To get a better idea of the model's performance in the real world, we further evaluate the performance of the fine-tuned 0.22B parameter CodeT5-base from the tool on real user data that was collected during the user studies. The distribution of NL queries collected from the user studies represents the ``true" distribution of queries that can be expected from users in a low-code setting. Out of the 218 samples that were collected, we found only one sample in which a user explicitly specified a hyper-parameter value in their query.

We therefore only compute the accuracy of the operation name generated rather than the entire operation invocation (as they would use default values anyway and so the scores remain the same except for that one sample).

Out of 218 query requests, the fine-tuned CodeT5-base model that was used in our tool answered 150 queries correctly, which would suggest an overall accuracy of 68.8\%.
However, 33 of these requests targeted actions that are not supported by the sklearn API, such as dropping a column (commonly the territory of the Pandas library). Disregarding such unsupported usage, \nltool answered 141 out of 185 queries correctly for an overall accuracy of \textbf{76.2\%}. For 33 additional samples, neither annotator could infer a reasonable ground truth since the prompt was unclear (e.g.: ``empty"). Leaving these out, i.e., when the prompt is both clear \emph{and} the operator is supported by the tool, \nltool was accurate in over \textbf{90\%} (137/152) of completions \changeappendix{(refer to Section F in appendix for additional results)}.

\subsection{User Study}
\change{We conducted a user study with 20 participants with varying levels of AI expertise to create AI pipelines using \tool across four tasks, replacing \nltool with a simple keyword search in half the tasks.} We collect and analyze data to investigate the following research questions:

\begin{itemize}[leftmargin=.37in,nolistsep]
  \item [RQ1:] How do \nltool and other features help participants discover previously-unknown operators?
  \item [RQ2:] Are participants able to compose and then iteratively refine AI pipelines in our tool?
  \item [RQ3:] What are the benefits and challenges of using low-code for~AI?
\end{itemize}

\subsubsection{Study Methodology}
We recruited 20 participants within the same large technology company via internal messaging channels. \change{We expect that citizen developers without formal programming training may also have 
varying levels of AI expertise and intentionally solicited participants of all backgrounds. } 
Potential participants filled out a short pre-study survey to self-report experience in the following: machine learning, data preprocessing, and sklearn using a 1~(no experience) to 5~(expert) scale. Participants include a mix of roles including developers, data scientists, and product managers working in a variety of domains such as AI, business informatics, quantum computing, and software services.
25\% of the participants are female and the remaining 75\% are male. 
40\% of the participants self-reported being novices in machine learning by indicating a 1 or 2 in the pre-study survey. 

The study design is within-subjects~\cite{Creswell2013} where each participant was exposed to two conditions: using \tool with~(\emph{NL condition}) and without~(\emph{keyword condition}) the natural language~(NL) interface powered by 
\nltool.
The keyword condition used a simple substring filter for operator names. Each participant performed four tasks total across the two conditions. 
For each task, participants were instructed to create AI pipelines with data preprocessing and classifier steps on a sample dataset with as high a score~(accuracy on the test set) as possible during a time period of five to ten minutes. Each sample dataset was split beforehand into separate train and test sets.
Tasks were open-ended with no guidance on what preprocessing steps or classifiers should be used.

There were four sample datasets in total and each participant was exposed to all four. The sample datasets are public tabular datasets from the UCI Machine Learning Repository~\cite{ucidatasets}.
Two of the tasks (A~and~D) require a specific data preprocessing step in order to successfully create a pipeline while two (B~and~D) technically do not require preprocessing to proceed.
For each participant, the order of the conditions and the order of the tasks were shuffled such that there is a uniform distribution of the order of conditions and tasks.

As our study included machine learning novices, we gave each participant a short overview of the basics of machine learning with tabular datasets and data preprocessing. We avoided using specific terms or names of operators in favor of more general descriptions of data-related problems.

We then gave each participant an overview of \tool. To mitigate potential biasing or priming, the tool overview used a fifth dataset from the UCI repository~\cite{ucidatasets}.
To avoid operators that were potentially useful in user tasks, the overview used both a non-sklearn operator that was not available in the study versions of the tool as well as sklearn's DummyClassifier
that generates predictions without considering input features. 
Participants were allowed to use external resources such as web search engines or documentation pages. Nudges were given by the study administrators after five minutes if necessary to help participants progress in a task. Nudges were in the form of reminders to use tool features such as the NL interface, external resources, or to include missing steps such as data preprocessing or classifiers. Nudges did not mention specific operator names nor guidance on specific actions to take.

For each version of the tool, study administrators would describe the unique features of the particular version and then have participants perform tasks using two out of four sample datasets. After performing tasks using both versions of the tool and all four sample datasets, participants were asked to provide open-ended feedback and/or reactions for both \tool and the comparison between the NL and keyword conditions.

\begin{table}[t]
\begin{center}
\small
\caption{Incidence of tasks where participants find previously-unknown operators per condition (40 tasks for all, 16 tasks by novices, and 24 by non-novices). Note that rows may not sum to 100\% as participants can use multiple methods to discover operators for a given task or not discover operators at all.}
\vspace{-2mm}
\begin{tabular}{|r|r|r|r|r|}
    \hline
    \multirow{2}{*}{\textbf{Condition}} &  \multirow{2}{*}{\textbf{Participant}} & \multicolumn{3}{c|}{\textbf{Method of Discovery}} \\\cline{3-5}
     &  & \nltool & Web search & Palette \\
    \hline
    \multirow{3}{*}{NL} & All & 30 (75.0\%) & 5 (12.5\%) & 5 (12.5\%)\\
     & Novice & 8 (50.0\%) & 2 (12.5\%) & 4 (25.0\%) \\
     & Non-Novice & 22 (91.7\%) & 3 (12.5\%) & 1 \hspace*{3.5pt}(4.2\%)\\
    \hline
    \multirow{3}{*}{Keyword} & All &  \multirow{3}{*}{\parbox{1.5cm}{\emph{Not available in this condition.}}}& 13 (32.5\%) & 11 (27.5\%) \\
     & Novice &  & 3 (18.8\%) & 5 (31.3\%) \\
     & Non-Novice &  & 10 (41.7\%) & 6 (25.0\%) \\
    \hline
\end{tabular}
\vspace{-4mm}
\label{tab1:discoverunknown}
\end{center}
\end{table}

\subsubsection{Data Collection and Analysis}

To answer our research questions, for each participant, we collect and analyze both quantitative and qualitative data.
For quantitative data, we report on the incidence of participants discovering a previously-unknown operator~(RQ1) and the incidence of completing the task and iterating or improving the pipeline~(RQ2). We consider an operator `previously-unknown' if the participant found and used the operator without using the exact or similar name. For example, using an NL query such as \textit{``deal with missing values''} to find the \texttt{SimpleImputer} operator is considered discovering a previously-unknown operator while a query such as \textit{``simpleimpute''} is not. We report discovery using the following methods: through \nltool, generic web search engine~(Google), and scrolling through the palette. Participants may discover multiple unknown operators during the same task, possibly using different methods. 
For each participant's task, we consider it `complete' if the composed pipeline successfully trains against the dataset's training set and returns a score against the test set. We consider the pipeline iterated if a participant modifies an already-complete pipeline. More specifically, we consider the following forms of iteration: a preprocessing operator block is added or swapped, a classifier block is swapped, or hyper-parameters are tuned. We report each of these as separate types of pipeline iteration. Participants may perform multiple types of iteration during the same task. 
Both sets of quantitative metrics are counted per task~(80 tasks total for 20 participants, 40 tasks per condition).

We use qualitative data to answer RQ3. This data focuses on the participants' actions in \tool, commentary while using the tool and performing tasks, and answers to open-ended questions after the study.
Specifically, the same two authors that administered the user study analyzed the notes generated by the study along with the audio and screen recordings when the notes were insufficient, using discrete actions and/or quotations as the unit of analysis.
The first round of analysis performed open coding~\cite{Creswell2013} on data from 16 studies to elicit an initial set of 73 themes. The two authors then iteratively refined the initial themes through discussion along with identifying 13 axial codes which are summarized in Figure~\ref{fig:axial}. The same authors then performed the same coding process on a hold-out set of 4 studies. No additional themes were derived from the hold-out set of studies, suggesting saturation.

\subsubsection{Study Results}
\label{sec:user-study-results}

We answer RQ1 and RQ2 using quantitative data collected from observing participant actions per task and answer RQ3 through open coding of qualitative data.

\noindent\textbf{RQ1: How do \nltool and other features help participants discover previously-unknown operators?}

\Cref{tab1:discoverunknown} reports how often participants discovered previously-unknown operators during their tasks.
80\% of the participants discovered an unknown operator across 63.8\% of all 80 tasks in the study. 
\change{Participants discovered unknown operators in 82.5\% of the 40 NL condition tasks compared to 45\% of the 40 keyword condition tasks. The odds of discovering an unknown operator are significantly greater in the NL condition than keyword~($p\mlt0.001$) using Barnard's exact test. We examine the methods of discovery in more detail, noting that 
\nltool is only available in the NL condition whereas web search and scrolling through the operator palette are available in both conditions. We note that the participants were not able to use the keyword search to discover unknown operators due to needing at least part of the exact name.}
Using
\nltool, participants discovered unknown operators in \change{75\% of tasks in the NL condition as opposed to an average of 22.5\%  using web search engines (12.5\% in the NL condition and 32.5\% in the keyword condition) and an average of 20\% by scrolling through the operator palette (12.5\% in the NL condition and 27.5\% in the keyword condition).}
\change{Within the NL condition, the odds of an unknown operator being discovered are significantly greater using \nltool as opposed to both web search~($p\mlt0.001$) and scrolling~($p\mlt0.001$).}
When splitting on the experience of the participant, we find statistically greater chances of \change{novices discovering operators in the NL condition using \nltool as opposed to web search~(p=0.013) but not scrolling~(p=0.086). Non-novices were significantly more likely to discover operators using \nltool compared to web search or scrolling~($p\mlt0.001$, $p\mlt0.001$). 
Results do not change if considering web searches or scrolling across all 80 tasks.}
\change{These results suggest that \nltool is particularly helpful in discovering previously-unknown operators, especially compared to web search, but novices still face some challenges.}
We discuss these challenges in RQ3.

\noindent\textbf{RQ2: Are participants able to compose and then iteratively refine AI pipelines in our tool?}

\Cref{tab1:composeiterate} reports how often participants iterated on pipelines.
Participants completed 82.5\% of the 80 tasks in the study and further iterated their pipelines in 72.5\% of the tasks. Splitting on condition, the NL condition has 85\% task completion and 72.5\% further iteration while the keyword condition has 80\% task completion and 72.5\% iteration rate.
Swapping classifiers was the most common form of iteration at 48.8\%, followed by adding or swapping preprocessors at 43.8\% and setting hyper-parameters at 30\%. 
Comparing novices to non-novices, both types of participants are mostly successful in iterating pipelines with no significant differences in iteration rate using Barnard's exact test~(p=0.109).
This result holds when iterating preprocessors~(p=0.664) but not classifiers~(p=0.038) nor hyper-parameters~(p=0.005). Non-novices are more likely to complete the task than novices~(p=0.002).
Regardless of experience, both novices and non-novices are able to iteratively refine their pipelines, but novices face some challenges compared to non-novices regarding actually completing the task. These challenges are discussed in the next research question. 

\begin{table}[t]
    \begin{center}
    \small
    \caption{\label{tab1:composeiterate}Incidence of tasks where participants complete and iterate on preprocessors, classifiers, and hyper-parameters.}
    \vspace*{-2mm}
    \begin{tabular}{|r|c|c|c|}
    \hline
    \multirow{2}{*}{\textbf{Iteration Type}} & \multirow{2}{*}{\parbox{1.5cm}{\centering\textbf{\textit{Total Tasks}\\(80)}}} & \multirow{2}{*}{\parbox{1cm}{\centering\textbf{\textit{Novice}\\(32)}}} & \multirow{2}{*}{\parbox{1.5cm}{\centering\textbf{\textit{Non-Novice}\\(48)}}}\\
    & & &  \\
    \hline
    Task Completion        & 66 (82.5\%) & 21 (65.6\%) & 45 (93.8\%) \\
    \hline
    Swap Classifier        & 39 (48.8\%) & 11 (34.4\%) & 28 (58.3\%) \\
    Add/Swap Preprocessors & 35 (43.8\%) & 15 (46.9\%) & 20 (41.7\%) \\
    Set Hyper-parameters   & 24 (30.0\%) & \hspace*{3.5pt}4 (19.0\%) & 20 (41.7\%) \\
    All Iterations         & 58 (72.5\%) & 20 (62.5\%) & 38 (79.2\%) \\
    \hline
    \end{tabular}
    \vspace{-4mm}
    \end{center}

\end{table}

\noindent\textbf{RQ3: What are the benefits and challenges of using low-code for AI?}

\begin{figure}[t]
\centerline{\includegraphics[width=0.9\linewidth]{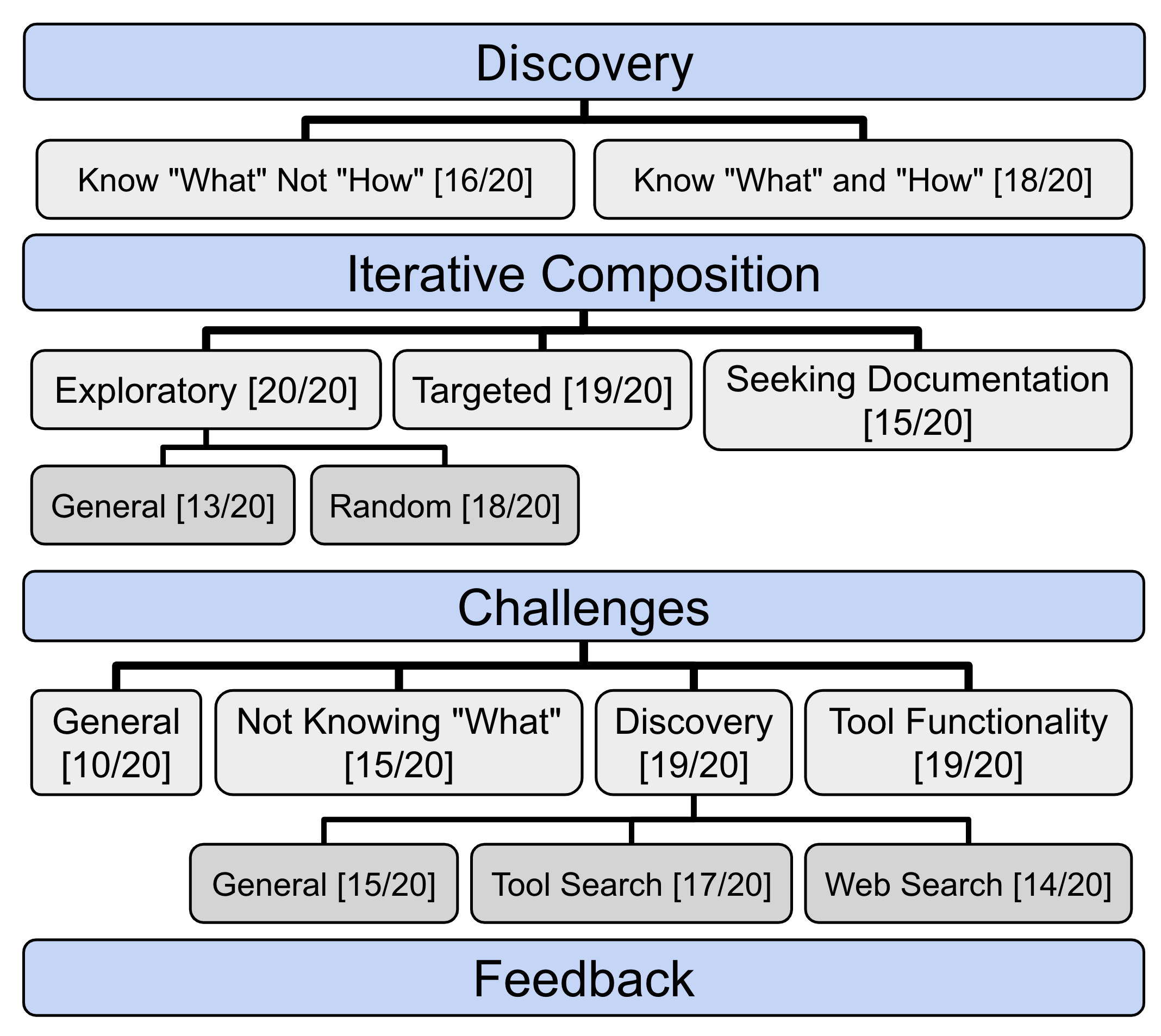}}
\vspace*{-4mm}
\caption{Axial codes from our qualitative analysis.}
\label{fig:axial}
\vspace*{-4mm}
\end{figure}

\Cref{fig:axial} shows our 13 axial codes for answering RQ3.
These codes broadly represent three overarching themes regarding working with low-code and machine learning:
1)~\textit{Discovery} of machine learning operators relevant for the task at hand,
2)~\textit{Iterative Composition} of the operators in the tool, and
3)~\textit{Challenges} that participants, particularly novices, face regarding working with machine learning and/or using low-code tools. We also collect \textit{Feedback} from participants to inform future development of \tool. Due to space limitations, we only report on a selection of the 13 axial codes and 73 codes derived from open coding \changeappendix{(refer to Section G in the appendix for the full list of codes)}.

For the first category of \textbf{Discovery}, our analysis derived two axial codes related to the participants' goal while attempting to discover operators: 1) \textit{Know ``What'' Not ``How''} where participants have a desired action in mind but do not know the exact operator that performs that action~(19 out of 20 participants experienced this axial code) and 2) \textit{Know ``What'' And ``How''} where participants have a particular action and operator in mind~(18/20). We dive deeper into \textit{Know ``What'' Not ``How''} which includes the code where participants \textit{Discover a previously-unknown operator using NL}~(16/20). We found in RQ1 that \nltool was helpful in finding unknown operators compared to other methods. The qualitative data suggests that participants were able to find unknown operators using \nltool during cases where they have an idea of the action to perform but do not know the exact operator name for a variety of reasons. For example, when discovering \texttt{SimpleImputer} with \nltool, P11 noted that they \emph{``never used SimpleImputer but had an idea of what I wanted to do, even though I generally remove NaNs in Pandas.''} Another example is P16 who \emph{``preferred the [NL version of \tool], even when I was doing Google searches, they... didn’t give me options, your tool at least returns some options that I can try out and swap out.''}
As a novice, P16 had difficulties finding the names of useful operators from web search results as opposed to the \nltool which directly returned actionable operators. We note that challenges regarding general web search is also an axial code.

For the second category of \textbf{Iterative Composition}, we derived four axial codes related to participant behaviors while attempting to compose and iterate on pipelines: 
1) \textit{General Exploratory}~(13/20) iteration, 2) Exploratory iteration but where participants will select operators or hyper-parameters seemingly at \textit{Random}~(18/20), 3) \textit{Targeted}~(19/20) iteration where participants select operators or hyper-parameters with a particular intent, and 4) \textit{Seeking Documentation}~(15/20) where participants search for documentation to inform iteration decisions. We note that for both forms of Exploratory iteration and Targeted iteration, we find examples of participants iterating classifiers, preprocessors, and hyper-parameters. 
For the axial code of seemingly \textit{Random} iteration, participants, especially~(but not exclusively) novices, when unsure of how to proceed, tended to try out arbitrary preprocessors or classifiers. This was more common for more difficult tasks that required particular data preprocessing to proceed. For example, non-novice P9
remarked \emph{``I'm not familiar enough with it, so do I Google it or brute force it? [...] I don't even know what to Google to figure this out... I guess I'll do some light brute-forcing''} and proceeded to swap in and out preprocessors from the palette. 
In contrast, the axial code of \textit{Targeted}~(19/20) iteration has codes that reflect particular intentions that participants derived from observations within the tool, such as \textit{Noticing error messages}~(10/20) or \textit{Making use of data tables in task}~(14/20). As an example of the data tables case, P11 realized through the \textit{Before} data table that the given dataset had \emph{``too many columns''} and added the \texttt{IncrementalPCA} operator along with setting its \texttt{n\_components} hyper-parameter to 5. Upon seeing the change in data in the \textit{After} data table, they remarked, \emph{``Wow... I really like that I can see all the hyper-parameters that I can play with''} and proceeded to tune various hyper-parameters.

The third category is the variety of \textbf{Challenges} that participants faced while using \tool and performing the machine learning tasks where we derive six axial codes: 1) \textit{General} challenges~(10/20) faced by participants that are not particular to our tool or tasks, 2) \textit{Not Knowing ``What''}~(15/20) where participants experienced difficulties due to knowing neither ``what'' nor ``how'' to begin, 3) \textit{General Discovery} challenges~(15/20), 4) Discovery challenges around using \textit{Web search}~(14/20), 5) Discovery challenges when using \textit{Tool search}~(17/20) or specifically using \nltool, and 6) \textit{Tool Functionality}~(19/20) which describes challenges participants faced using (or not using) \tool features. 
We dive deeper into the axial code of \textit{Not Knowing ``What''} and note its contrast to the \textit{Know ``What'' Not ``How''} axial code where participants may have intentions but not know how to execute them or the \textit{Exploratory} iteration axial code where participants may not have specific intentions but know how to iterate. 
All novices~(8/8) and most non-novices~(7/12) experienced this challenge. 
The primary code is that participants \textit{Did not know ``what'' they wanted to do}~(11/20). One possible cause of this lack of progression is choice paralysis, for example on P17's first task, \emph{``first things first, I don't even know where to begin... right now it's super overwhelming, I guess I'll start throwing stuff in there.''}
We also describe the axial code of \textit{Tool search}~(17/20) where participants had difficulties forming search queries for \nltool.
 
Participants noted that despite the interface being intended for general natural language, the interface still \textit{Needed a specific vocabulary}~(8/20). 
As P19, a novice, described it, \emph{``I get the idea of how it's supposed to work but it's hit and miss... even if I use very layman's terms... it expects a non-naive explanation of what needs to be done.'' }
Part of this challenge may be due to a mismatch in the natural language in Kaggle notebooks used to train \nltool and the language used by novices. 

\section{Discussion}
Our results show that both the \vptool and \nltool components were helpful with aspects like operator discovery~(RQ1) or iteratively composing pipelines~(RQ2), even for novice participants. This is useful for \change{citizen developers who have an idea of \emph{what} they would like to do but do not fully know \emph{how} to accomplish that, perhaps due to a lack of formal programming training}. In fact, our qualitative analysis~(RQ3) reveals that a number of our participants~(including all novices who participated) struggled with knowing \emph{what} to do.
\change{End-users writing software face similar ``design barriers''~\cite{ko_myers_aung_2004}, where it is difficult for a non-programmer to even conceptualize a solution.}
In contrast to other popular low-code domains such as traditional software~\cite{resnick_et_al_2009}, the domain of developing machine learning systems is particularly difficult in this regard due to its experimental nature where progress has a high degree of uncertainty~\cite{Wan2019}. This uncertainty then requires an abundance of judgment calls that rely heavily on prior machine learning experience~\cite{hill2016} that novices lack. Some participants in our studies echo this, identifying that some ML knowledge is necessary to use our tool\change{. That suggests that citizen developers who have some data science knowledge but lack programming training, such as statisticians, may benefit the most from our tool.}
A further improved low-code machine learning tool could thus be made more suitable towards~\change{novice} citizen developers by guiding them to discover the \emph{what} along with the \emph{how}, i.e., by helping developers acquire the necessary ML knowledge.

A potential extension, offered by a study participant, is to provide suggestions in the form of templates or recipes for pipelines. These suggestions could also be contextual to the given dataset or active pipeline, for example automatically suggesting encoders when detecting categorical features. \change{Ko et al.~\cite{ko_myers_aung_2004} also suggest templates as a possible solution for design barriers.} A related suggestion made by a number of our study participants is to implement data visualization and summarization tools for the given dataset, such as plots, charts, confusion matrices, etc. These visualizations could themselves inform contextual suggestions -- a histogram detecting a non-standard distribution may suggest the need for a \texttt{StandardScaler}. These contextual suggestions may also help in guiding developers in \emph{what} to do, making for a more generally useful low-code tool for both \change{citizen and experienced developers alike.} 

\textbf{Threats to Validity:}
The user study for \tool has several limitations. The study focused on relatively small, public tabular datasets and scikit-learn operators and may not be indicative of other machine learning tasks such as deep learning on large datasets. Participants also all come from the same large technology company and may not be representative of general users. However, we did intentionally elicit participation from a variety of groups and experience levels to mitigate this. As our user study has a within-subjects design, there may be potential learning effects between tasks and conditions. In fact, we observed some cases of this (8/20), with some participants explicitly mentioning selecting particular operators due to the previous task. We mitigated this learning effect by randomizing the order of tasks and conditions, as well as by having two tasks (A~and~D) require the use of preprocessing operators that were not applicable to other tasks. 

\section{Conclusion}
We developed \tool, a low-code tool that combines visual programming, \vptool, and programming by natural language (PBNL), \nltool, to help developers of all backgrounds create AI pipelines. 
We used \tool to provide some of the first insights into whether and how visual programming and PBNL help programmers by conducting user studies across four tasks with (NL condition) and without (keyword condition) \nltool.
Overall, \tool helped developers compose (85\% of tasks) and iterate (72.5\% of tasks) over AI pipelines. Furthermore, \nltool helped users discover previously-unknown operators in 75\% of tasks, compared to just 22.5\% (12.5\% in the NL condition and 32.5\% in the keyword condition) using web search. 
Our qualitative analysis showed that PBNL helped users discover \textit{how} to implement various parts of the pipeline when they know \textit{what} to do. However, it failed to support novices when they lacked clarity on what they want to accomplish, which may suggest a worthwhile target for improving AI-based program assistants.
Our work demonstrates the promise of combining both an AI-powered natural language interface and a visual interface for helping developers of all backgrounds create AI pipelines without writing code.

\section{Data Availability}
The implementation of \tool, datasets for training and evaluating \nltool, results of additional experiments, as well as the material from the user study, incl. the full set of (axial) codes \& anonymized quantitative and qualitative data are available at: \url{https://doi.org/10.5281/zenodo.7042296}.

\bibliographystyle{ACM-Reference-Format}
\bibliography{references}

\begin{appendices}


\section{ Task formulations}

\Cref{tab:example-tasks} contains additional examples of NL queries and the corresponding code based on the task formulation.

\begin{table*}[h]
  \begin{center}

    \caption{Examples of NL-Code pairs for different task formulations.}
    \small
    \label{tab:example-tasks}
    \vspace*{-2mm}
    \begin{tabular}{|p{2.5cm}|p{3.4cm}|p{3.4cm}|p{3.4cm}|p{3.4cm}|}

    \hline
    \multirow{2}{*}{\textbf{NL query}} & \multirow{2}{*}{\textbf{Operator Name}} & \textbf{Complete\newline Operator Invocation} & \textbf{Masked\newline Operator Invocation} & \textbf{Hybrid \newline Operator Invocation} \\
    \hline
    Split X, y data into training set and testing set & train\_test\_split & train\_test\_split(X, y,\newline test\_size=0.2) & train\_test\_split(MASK, MASK, test\_size=MASK) & train\_test\_split(X, y,\newline test\_size=MASK)\\
    \hline
    PCA with 2 components & PCA & PCA (n\_components=2, random\_state=42) & PCA (n\_components=MASK, random\_state=MASK) & PCA (n\_components=2, random\_state=MASK) \\
    \hline
    Replace missing data with the mean value & SimpleImputer & SimpleImputer\newline(strategy=`mean') & SimpleImputer\newline(strategy=MASK) & SimpleImputer\newline(strategy=`mean') \\
    \hline
    Encoding categorical features & OneHotEncoder & OneHotEncoder() & OneHotEncoder() & OneHotEncoder() \\
    \hline
    Standardisation of Data  & StandardScaler & StandardScaler() & StandardScaler() & StandardScaler() \\
    \hline
    K-Means with 4 clusters & KMeans & KMeans (n\_clusters=4, random\_state=42) & KMeans (n\_clusters=MASK, random\_state=MASK) & KMeans(n\_clusters=4,  random\_state=MASK) \\
    \hline
    Build Decision Tree with max\_depth = 7 & DecisionTreeClassifier & DecisionTreeClassifier (criterion=`gini', max\_depth=7) & DecisionTreeClassifier (criterion=MASK, max\_depth=MASK) & DecisionTreeClassifier (criterion=MASK, max\_depth=7) \\
    \hline
    Random forest with balanced class weight  & RandomForestClassifier & RandomForestClassifier (n\_estimators=100, class\_weight=`balanced') & RandomForestClassifier (n\_estimators=MASK, class\_weight=MASK) & RandomForestClassifier (n\_estimators=MASK, class\_weight=`balanced')\\
    \hline
    \end{tabular}
  \end{center}

\end{table*}

\section{Additional details about the data}
\label{sec:data-parameters}
We find that the NL query corresponds to a single sklearn operator invocation in the code cell in 62\% of the data, whereas in the remaining 38\% it has multiple sklearn operator invocation statements. 
\Cref{tab:params} shows the distribution of hyper-parameters in hybrid operator invocations, based on whether the hyper-parameters were named, and whether the hyper-parameter values were masked or valued (based on whether they appear in the NL query).

\begin{table}[h]
\small
  \begin{center}
    \caption{Distribution of hyper-parameters in hybrid operator invocations.}
    \vspace*{-2mm}
    \label{tab:params}
    \begin{tabular}{|P{2.5cm}|P{1.5cm}|P{1.5cm}|P{1.5cm}|}

    \hline
     \multirow{2}{*}{\textbf{Parameter Type}} & \multicolumn{3}{|c|}{\textbf{Distribution of Parameters (\%)}} \\
    \cline{2-4}
     & \textbf{0} & \textbf{1-3} & \textbf{4+} \\
    \hline
    Total & 18.49 & 61.51 & 19.99\\
    \hline
    Named & 54.82 & 39.16 & 6.01\\
    \hline
    Masked & 18.99 & 61.61 &19.39 \\
    \hline
    Valued & 96.97 & 3.02 & 0.01 \\
    \hline
    \end{tabular}
  \end{center}
\end{table}

\section{Decoding Techniques}
\label{sec:decoders}
We experiment with different decoding techniques to generate our output hybrid operation invocation sequence. We describe them below:

\begin{itemize}[leftmargin=.37in,nolistsep]
\item[(i)]\textbf{Greedy decoding: }
At each time step, greedy decoding chooses the token with the highest conditional probability. Since the model weights are fixed, the output is deterministic, always yielding the same generation for a given prompt. We use this to generate a single output sequence.

\item[(ii)] \textbf{Top K sampling: }
At each time step, only consider the top $k$ most probable tokens (according to the model). Renormalize their probabilities and select a token corresponding to these probabilities. The output of this approach is no longer deterministic, but instead explores multiple, predominantly high-probability, completion paths. For our experiments, we set the value of $k$ to 5 and generate a total of 5 output sequences.

\item[(iii)] \textbf{Nucleus sampling: }
Nucleus sampling is similar to top $k$ sampling, but rather than fixing the number of most-probable tokens to consider at each time step, it samples a variable number of tokens whose cumulative conditional probabilities reaches or exceeds a defined probability ($p$) value.\footnote{This approach is also called ``top-$p$ sampling".} In contexts where just one or two tokens are highly probable or where many tokens are similarly plausible, this allows the model to switch between sampling more greedily or uniformly respectively. Once the samples are chosen, the probabilities are again redistributed among them and a token is selected according to these probabilities. In our experiments, we set the $p$ value to be 0.9 and generate a total of 5 output sequences.
\end{itemize}

\section{Modeling Results}
\label{sec:modeling-results-all}
We benchmark the performance of all the models and record the inference time across several different variables in \Cref{tab:model-results}, namely: 
\begin{itemize}[leftmargin=.37in,nolistsep]
\item[(i)] \textbf{Learning strategy:} we look at three different learning strategies for training the models, which include: training a sequence to sequence Transformer from scratch, fine-tuning a CodeT5 model, and few-shot prompting CodeGen, a large language model.
\item[(ii)] \textbf{Model size:} we compare a range of sizes for both the CodeT5 and CodeGen models. 
\item[(iii)] \textbf{Decoding:} we compare the different decoding strategies that include, greedy (top-1), topk (top-5) and nucleus (top-5).
\item[(iv)] \textbf{Tool supported operations:} Recall that since sklearn pipelines can only contain operators but not functions, \tool only exposes blocks for operators. Therefore, we also contrast the performance of the model on all the test data points (total of 7,351 samples) along with a filtered set of test data points that only include operators supported by the tool (resulting in 3,941 samples).
\end{itemize}

\begin{table*}[!ht]
\small

\caption{Accuracy scores for Hybrid Operator Invocation task across different model variations (*due to memory constraints)}
    \label{tab:model-results}
\vspace*{-2mm}
\begin{center}
    \begin{tabular}{|c|c|c|c|c|c|c|c|}
        \hline
         \multirow{4}{*}{\textbf{Model}} & \multirow{4}{*}{\textbf{Size}} & \multirow{4}{*}{\textbf{Decoding}} &  \multicolumn{5}{c|}{\textbf{Test notebooks}} \\\cline{4-8}
    
         & & & \multirow{3}{*}{\textbf{Time}} & \multicolumn{4}{c|}{\textbf{Accuracy in k (\%)}} \\\cline{5-8}
        
        & & & & \multicolumn{2}{c|}{\textbf{OpName}} & \multicolumn{2}{c|}{\textbf{OpInvocation}} \\\cline{5-8}
        
        & & & \textbf{(s)} & \textbf{all} & \textbf{tool} & \textbf{all} & \textbf{tool} \\
        
        \hline
        \multirow{3}{*}{Transformer} & \multirow{3}{*} {6 layers} & greedy (n=1) & 0.20 & 43.23 & 47.39 & 16.81 & 14.13 \\
        \cline{3-8}
         & & topK (n=5) & 1.11 & 60.13 & 63.74 & 29.16 & 28.21 \\
        \cline{3-8}
         & & nucleus (n=5) & 1.21 & 59.62 & 62.47 & 29.99 & 27.88 \\
        \hline
        \multirow{9}{*}{Fine-tuned CodeT5} & \multirow{3}{*} {small} & greedy (n=1) & 0.76 & 57.43 & 64.19 & 23.88 & 21.44 \\
        \cline{3-8}
         & & topK (n=5) & 1.09 & 71.78 & 77.01 & 38.86 & 39.25 \\
        \cline{3-8}
         & & nucleus (n=5) & 1.52 & 71.20 & 76.50 & 39.57 & 39.81 \\
        \cline{2-8}
         & \multirow{3}{*} {base} & greedy (n=1) & 1.55 & 59.37 & 65.66 & 25.84 & 23.97 \\
        \cline{3-8}
         & & topK (n=5) & 2.37 & 73.20 & 78.07 & 40.04 & 39.93 \\
        \cline{3-8}
         & & nucleus (n=5) & 3.15 & 73.35 & 77.94 & 41.19 & 40.97 \\
        \cline{2-8}
         & \multirow{3}{*} {large} & greedy (n=1) & 3.66 & 60.01 & 64.96 & 26.90 & 24.56 \\
        \cline{3-8}
         & & topK (n=5) & 5.89 & 73.57 & 77.47 & 41.27 & 40.73 \\
        \cline{3-8}
         & & nucleus (n=5) & 6.50 & 73.22 & 77.16 & 40.27 & 39.63 \\
        \hline
        \multirow{12}{*}{CodeGen} & \multirow{3}{*} {350M} & greedy (n=1) & 4.39 & 19.65 & 25.09 & 1.94 & 3.35\\
        \cline{3-8}
         & & topK (n=5) & 5.05 & 46.02 & 57.59 & 9.66 & 15.25\\
        \cline{3-8}
         & & nucleus (n=5) & 5.42 & 43.85 & 55.16 & 9.65 & 15.19\\
        \cline{2-8}
         & \multirow{3}{*} {2.7B} & greedy (n=1) & 6.52 & 22.67 & 30.55 & 2.41 & 4.21 \\
        \cline{3-8}
         & & topK (n=5) & 8.59 & 48.63 & 60.62 & 9.48 & 15.81 \\
         \cline{3-8}
         & & nucleus (n=5) & 9.56 & 47.33 & 59.27 & 9.31 & 15.20 \\
        \cline{2-8}
         & \multirow{2}{*} {6.1B} & greedy (n=1) & 8.09 & 26.01 & 35.70 & 3.51 & 6.06 \\
        \cline{3-8}
         & & topK (n=5) & 10.24 & 49.79 & 61.94 & 11.32 & 17.94 \\
        \cline{3-8}
         & & nucleus (n=5) & 10.43 & 48.35 & 60.39 & 11.03 & 17.20 \\
         \cline{2-8}
         & \multirow{2}{*} {16.1B} & greedy (n=1) & 10.99 & 24.60 & 34.66 & 3.44 & 5.40 \\
        \cline{3-8}
         & & topK (n=3*) & 14.27 & 43.27 & 55.41 & 9.73 & 13.98 \\
         \cline{3-8}
         & & nucleus (n=3*) & 14.25 & 41.34 & 53.89 & 10.01 & 14.62\\
        
        \hline
        \end{tabular}
        
\end{center}
\end{table*}

We also perform additional experiments in an effort to reduce the inference time of the fine-tuned CodeT5 model with top-5 decoding for integration with the tool. We reduce the output sequence length from 512 tokens to 64 and find a negligible decline in accuracy~(73.20\% for 512 tokens vs.\ 72.81\% for 64 tokens) with significantly lower inference time~(2.37s vs.\ 1.56s per sample). Note that the inference time is not proportional to the number of tokens as the model learns to stop generating new tokens when it hits the end token.

\section{Annotation guidelines for real user data}
\label{real-query-annotation}

Two authors manually annotate the real user data by looking at the NL query and the predictions returned by \nltool. More specifically, we look at the following criteria during annotation (multiple annotations are possible per query):

\begin{itemize}[leftmargin=.37in,nolistsep]
    \item Accurate prediction: At least one of the predictions returned by the model matches the inferred user intent in the query. Inferred intent was determined by annotator domain knowledge.
    \item NL unclear: The NL query is unclear (e.g. \textit{``empty''}, \textit{``numbers''}, \textit{``variable''} ) when neither annotator could infer the intent from the query alone. 
    \item Partially correct: The predictions are partially correct. This usually happened if the NL query requests multiple operators or intents such as \textit{``normalize features and run linear regression''}. 
    \item Not supported by tool: The NL query requests targeted actions that are not supported by the sklearn.
    \item No output returned by model: These are the cases where the model fails to return any usable predictions.
\end{itemize}

\Cref{tab:annotation} has the distribution of data per task for all the different properties we look at when manually annotating the real user data.

\begin{table*}[h]
\small
  \begin{center}
    \caption{Distribution of various properties annotated manually for real user data. }
    \vspace*{-2mm}
    \label{tab:annotation}
    \begin{tabular}{|c|c|c|c|c|c|c|}

    \hline
    Task & Total & Accurate prediction & NL unclear & Partially correct & Not supported by tool & No output returned \\
    \hline
    A & 43 & 30 & 5 & 3 & 3 & 0\\
    \hline
    B & 41 & 27 & 4 & 7 & 16 & 3\\
    \hline
    C & 53 & 38 & 9 & 3 & 6 & 4\\
    \hline
    D & 81 & 55 & 22 & 11 & 8 & 6 \\
    \hline
    All & 218 & 150 & 40 & 24 & 33 & 13\\
    \hline
    \end{tabular}
  \end{center}
\end{table*}

\section{Model evaluation on real user data}
\label{real-eval}

We evaluate the performance of \nltool on the annotated real user data. Here is a summary of the findings:
\begin{itemize}[leftmargin=.37in,nolistsep]
    \item Total accuracy = 150/218 = 68.80\%
    \item Percentage of data where the NL query is clear = 178/218 = 81.65\%
    \item Percentage of data where the NL query is not clear = 40/218 = 18.35\%
    \item Accuracy of model when the NL query is clear = 145/178 = 81.46\%
    \item Accuracy of model when the operator is supported by tool = 141/185 = 70.81\%
    \item Accuracy of model when the NL query is clear and operator is supported by tool = 137/152 = 90.13\%
\end{itemize}

\section{User Study}

\subsection{User Study Task Datasets}

The sample datasets used in the user study task from the UCI Repository are as follows:
\begin{itemize}[leftmargin=.37in,nolistsep]
    \item Dataset~A modifies the Iris dataset to include missing values in the form of not-a-number~(NaN) for 30\% of values. 
    \item Dataset~B is the Covertype dataset and demonstrates features with differing scales.
    \item  Dataset~C is the Digits dataset and demonstrates relatively higher dimensionality.
    \item Dataset~D is a modified version of the Mushroom dataset that only contains categorical features.
    \item The tutorial uses the Abalone dataset. 
\end{itemize}
While Datasets A and D require a specific data preprocessing step in order to successfully create a pipeline, B and D do not technically require preprocessing to proceed. The specific datasets and train/test splits used are also available as artifacts.

\subsection{Results}

The following is the full listing of codes and axial codes from the qualitative analysis, broken down by high-level category (which corresponds to 1st level axial codes): 1) Discovery~(\Cref{tab:codes-discovery}), Iterative Composition~(\Cref{tab:codes-ic}), Challenges~(\Cref{tab:codes-challenges}), and Feedback~(\Cref{tab:codes-feedback}).

\begin{table*}[!ht]
\small

\caption{Full codes and axial codes from qualitative analysis for Discovery category.}
\label{tab:codes-discovery}
\resizebox{\textwidth}{!}{%
\begin{tabular}{|l | l | l | r |}
\hline
\textbf{1st Level Axial}  & \textbf{2nd Level Axial}    & \textbf{Code}    & \textbf{Participant Count (20)} \\
\hline
Discovery             & Know "What" Not "How"                & Discovered operator they didn't know about using NL                     & 16          \\
Discovery             & Know "What" Not "How"                & Discovered useful operator by browsing toolbox                          & 7           \\
Discovery             & Know "What" Not "How"                & Google error message to find solution                                   & 1           \\
Discovery             & Know "What" Not "How"                & Google for how to do something find general term to use in tool         & 13          \\
Discovery             & Know "What" Not "How"                & Keyword result does not match what they want                            & 5           \\
Discovery             & Know "What" Not "How"                & Liked NLP version                                                       & 13          \\
Discovery             & Know "What" Not "How"                & NL search using ML terms                                                & 6           \\
Discovery             & Know "What" Not "How"                & Scrolling through toolbox for something they recognize or relevant name & 10          \\
Discovery             & Know "What" Not "How"                & Search same as hint                                                     & 1           \\
Discovery             & Know "What" and "How"                & Keyword close to operator name but not exact match                      & 6           \\
Discovery             & Know "What" and "How"                & Keyword search using ML term                                            & 10          \\
Discovery             & Know "What" and "How"                & Searched for exact operator name                                        & 14          \\
\hline
\end{tabular}%
}
\end{table*}

\begin{table*}[!ht]
\small
\caption{Full codes and axial codes from qualitative analysis for Iterative Composition category.}
\label{tab:codes-ic}
\resizebox{\textwidth}{!}{%
\begin{tabular}{|l|l|l|l|r|}
\hline
\textbf{1st Level Axial}  & \textbf{2nd Level Axial} & \textbf{3rd Level Axial}   & \textbf{Code}    & \textbf{Participant Count (20)} \\
\hline
Iterative Composition & Exploratory           & General       & Liked visual blocks                                                     & 5           \\
Iterative Composition & Exploratory           & General       & Used score to determine how to refine                                   & 14          \\
Iterative Composition & Exploratory           & Random        & Randomly refining data processing                                       & 9           \\
Iterative Composition & Exploratory           & Random        & Randomly refining hyperparameters                                       & 4           \\
Iterative Composition & Exploratory           & Random        & Randomly refining model                                                 & 14          \\
Iterative Composition & Exploratory           & Random        & Randomly scroll through toolbox                                         & 17          \\
Iterative Composition & Exploratory           & Random        & Searched for exact operator name                                        & 1           \\
Iterative Composition & Exploratory           & Random        & Unclear goal but pipeline worked                                        & 4           \\
Iterative Composition & Targeted              &               & Google for hyperparameter values                                        & 4           \\
Iterative Composition & Targeted              &               & Intentionally refine model                                              & 5           \\
Iterative Composition & Targeted              &               & Intentionally refine preprocessing                                      & 5           \\
Iterative Composition & Targeted              &               & Intentionally tuning hyperparameters                                    & 10          \\
Iterative Composition & Targeted              &               & Liked hyperparameter pane                                               & 2           \\
Iterative Composition & Targeted              &               & Made use of data tables in task                                         & 14          \\
Iterative Composition & Targeted              &               & Noticed error message                                                   & 10          \\
Iterative Composition & Targeted              &               & Used canvas to store blocks                                             & 4           \\
Iterative Composition & Targeted              &               & Used score to determine how to refine                                   & 1            \\
Iterative Composition & Seeking Documentation &               & Google for operator documentation                                       & 6           \\
Iterative Composition & Seeking Documentation &               & Hard to figure out what operator does                                   & 10          \\
Iterative Composition & Seeking Documentation &               & Hover over hyperparameters to learn more                                & 6           \\
\hline
\end{tabular}%
}
\end{table*}

\begin{table*}[!ht]
\small
\caption{Full codes and axial codes from qualitative analysis for Challenges category.}
\label{tab:codes-challenges}
\resizebox{\textwidth}{!}{%
\begin{tabular}{|l|l|l|l|r|}
\hline
\textbf{1st Level Axial}  & \textbf{2nd Level Axial} & \textbf{3rd Level Axial}   & \textbf{Code}    & \textbf{Participant Count (20)} \\
\hline
Challenges            & General               &               & Gave up on task                                                         & 3           \\
Challenges            & General               &               & Ignoring/misinterpreting error messages                                 & 4           \\
Challenges            & General               &               & Needed nudge to do something                                            & 8           \\
Challenges            & General               &               & Task clarification                                                      & 3           \\
Challenges            & Not Knowing "What"    &               & Did not know “what” they wanted to do                                   & 11          \\
Challenges            & Not Knowing "What"    &               & Learning effect - used exact same operator/pipeline as previous task    & 8           \\
Challenges            & Not Knowing "What"    &               & Nudge to prevent invalid pipeline                                       & 2           \\
Challenges            & Not Knowing "What"    &               & Tool easier to use when knowing "what" to do                            & 3           \\
Challenges            & Not Knowing "What"    &               & Unfamiliar with aspect of scikit-learn or machine learning              & 4           \\
Challenges            & Not Knowing "What"    &               & Used operator that did not work as intended                             & 4           \\
Challenges            & Discovery             & General       & Knew "what" they wanted to do but did not know the term                 & 15          \\
Challenges            & Discovery             & General       & Overwhelmed by choices                                                  & 2           \\
Challenges            & Discovery             & Google Search & Chose not to Google search                                              & 3           \\
Challenges            & Discovery             & Google Search & Google didn't help find sklearn operator (found pandas/numpy solution)  & 5           \\
Challenges            & Discovery             & Google Search & Google something at a very high level "classification/preprocessing"    & 6           \\
Challenges            & Discovery             & Google Search & Google something but not able to parse results                          & 10          \\
Challenges            & Discovery             & Google Search & Had difficulty articulating Google search                               & 3           \\
Challenges            & Discovery             & Tool Search   & Had difficulty articulating search inside tool                          & 5           \\
Challenges            & Discovery             & Tool Search   & NL returned unsupported operator                                        & 9           \\
Challenges            & Discovery             & Tool Search   & NL search is unclear or vague                                           & 10          \\
Challenges            & Discovery             & Tool Search   & Needed specific vocabulary                                              & 8           \\
Challenges            & Discovery             & Tool Search   & No results from keyword filter                                          & 9           \\
Challenges            & Discovery             & Tool Search   & No results returned from NLP                                            & 3           \\
Challenges            & Tool Functionality    &               & Did not understand aspect of tool                                       & 5           \\
Challenges            & Tool Functionality    &               & Didn’t use certain tool features                                        & 6           \\
Challenges            & Tool Functionality    &               & Learning curve required for tool                                        & 3           \\
Challenges            & Tool Functionality    &               & NLP returned wrong results                                              & 8           \\
Challenges            & Tool Functionality    &               & Non-deterministic results from NL search                                & 2           \\
Challenges            & Tool Functionality    &               & Non-deterministic scoring                                               & 4           \\
Challenges            & Tool Functionality    &               & Problem with tool                                                       & 4           \\
Challenges            & Tool Functionality    &               & They didn't understand what NL search really did                        & 5           \\
Challenges            & Tool Functionality    &               & Wanted to do something tool doesn't support                             & 11          \\
\hline
\end{tabular}%
}
\end{table*}

\begin{table*}[!ht]
\small
\caption{Full codes and axial codes from qualitative analysis for Feedback category.}
    \label{tab:codes-feedback}
\begin{tabular}{|l|l|r|}
\hline
\textbf{1st Level Axial}  & \textbf{Code}    & \textbf{Participant Count (20)} \\
\hline
Feedback                        & Comparison with other tool                                              & 4           \\
Feedback                       & Did not like NLP version                                                & 3           \\
Feedback                      & Did not like some feature of the tool                                   & 6           \\
Feedback                      & Do not like keyword version                                             & 2           \\
Feedback                          & Liked data table feature                                                & 8           \\
Feedback                          & Liked feature of tool                                                   & 12          \\
Feedback                     & Liked keyword version                                                   & 6           \\
Feedback                     & No preference between keyword and NLP                                   & 2           \\
Feedback                       & Suggestion for tool                                                     & 13          \\
Feedback                      & Wanted both NL and keyword                                              & 3           \\
\hline
\end{tabular}
\end{table*}

\end{appendices}

\end{document}